\documentclass[preprint,12pt]{elsarticle}

\usepackage{amsmath,amssymb,bm}
\usepackage{graphicx}
\usepackage{booktabs}
\usepackage{siunitx}
\usepackage{xcolor}
\usepackage{hyperref}
\usepackage{lineno}
\usepackage{caption}
\usepackage{subcaption}

\hypersetup{
    colorlinks=true,
    linkcolor=blue,
    citecolor=blue,
    urlcolor=blue
}

\begin{document}

\begin{frontmatter}

\title{Stochastic first-passage modeling of single-event burnout in SiC power MOSFETs}

\author[addr1]{Feiyi Liu}
\author[addr1]{Min Guo\corref{cor1}}
\ead{guomin1@cxtc.edu.cn}
\author[addr2]{Shiyang Chen}
\author[addr1]{Yuhan Jiang}
\author[addr1]{Mingyang Liu}
\author[addr3]{Yang Wang}

\cortext[cor1]{Corresponding author.}

\address[addr1]{School of Physics, Electrical and Energy Engineering, Chuxiong Normal University, Chuxiong 675000, China}
\address[addr2]{Department of Physics, Swansea University, SA2 8PP, Swansea, United Kingdom}
\address[addr3]{School of Information Science and Engineering, Shandong Institute of Petroleum and Chemical Technology, Dongying 257061, China}

\begin{abstract}
Single-event burnout (SEB) in silicon carbide (SiC) power MOSFETs is often characterized by deterministic threshold quantities. Near the
boundary between recovery and runaway, stochastic variability can make this threshold description probabilistic rather than sharp. This work introduces a first-passage perspective for stochastic threshold broadening in burnout. The process is described by a reduced electrothermal feedback-relaxation model with an absorbing boundary. The model combines carrier multiplication, avalanche feedback, localized heating, carrier loss, and thermal relaxation. Stochastic carrier and thermal terms represent unresolved event-level variability. The main finding is that finite fluctuations broaden the deterministic burnout threshold into a probabilistic transition band. Noise-induced subthreshold runaway also emerges, where nominally recoverable conditions can still fail through rare stochastic excursions. First-passage-time distributions resolve the time scale of burnout and survival probabilities further distinguish rapid feedback-dominated runaway from delayed stochastic failure. A feedback-relaxation phase diagram organizes recoverable, probabilistic, and rapidly unstable regimes. This framework provides a statistical-physics interpretation of threshold dispersion in single-event burnout of SiC power MOSFETs by linking coarse-grained electrothermal dynamics to probabilistic and time-resolved failure observables.
\end{abstract}


\end{frontmatter}

%\linenumbers

\section{Introduction}
\label{introduction}

Threshold-activated failure in nonlinear systems governed by external driving and dissipation is a central problem in statistical physics,
materials degradation, and electronic reliability \cite{Hanggi1990, Gabriel2023, Padovani2024}.
In these systems, a localized perturbation may relax back to a stable state if dissipation dominates the subsequent evolution, but it may also
be amplified into a runaway event if positive feedback exceeds the available relaxation capacity \cite{Zapperi1999,Spang2024, Fu2024}. 
A deterministic description often represents the transition from recovery to failure by a sharp threshold\cite{Kuehn2011,Shoji2015,Peng2021}. 
Realistic systems, however, contain microscopic fluctuations, event-to-event variability, and incomplete knowledge of local initial conditions 
conditions \cite{Gillespie1977,Weller2003,Schrimpf2007,Ham2024}. 
Nominally identical control parameters can therefore produce different trajectories near the critical region \cite{Reed2007}.
Failure is more naturally described by the probability that a stochastic trajectory reaches an absorbing boundary \cite{Redner2023, Schuss2007}. 
The corresponding statistics of survival and first passage provide a direct connection between threshold dispersion, rare subthreshold events, and stochastic-process theory \cite{Gardiner2009,VanKampen2007,Risken1996,Redner2001}.

Single-event burnout (SEB) in SiC power devices is a representative example of this class of nonlinear threshold phenomena \cite{Shoji2015,McPherson2020,Grome2024}. 
SiC power metal-oxide-semiconductor field-effect transistors (MOSFETs) have become important candidates for high-voltage and high-temperature power applications \cite{Baliga2008,Kimoto2014}. 
Their advantages originate from the wide band gap, high critical electric field, and good thermal properties of SiC. 
Under off-state bias, an energetic ion can generate a dense carrier track near a high-field region of the device \cite{Wang2024,Li2025}.
The generated carriers are swept by the electric field, amplified by impact ionization, and accompaniedby localized Joule heating through electrothermal work \cite{Zhang2022,Mo2023}. 
If carrier multiplication and heat generation reinforce each other faster than carrier extraction and thermal diffusion can relax the perturbation,
irreversible electrothermal runaway can occur \cite{Mo2023, Sexton2003, Pocaterra2023}. 
SEB is therefore both a reliability issue and a failure process governed by feedback, relaxation, and approach to a
failure boundary \cite{Grome2024,Sexton2003}.

The physical mechanisms of SEB have been studied since early investigations of power MOSFET failure under energetic-particle irradiation \cite{Sexton2003, Kuboyama1992}. 
Recent experimental, simulation, and review studies have linked SEB in SiC power devices to ion-induced current filamentation, avalanche amplification, localized power dissipation, and electrothermal runaway
\cite{Grome2024, Mo2023, Pocaterra2023}. 
These results indicate that burnout is governed not by primary ionization alone, but by coupled
electrothermal dynamics.
For SiC power MOSFETs, irradiation studies and technology computer-aided design (TCAD) simulations have provided a device-specific picture of electrothermal burnout. 
Ion-induced field perturbation and avalanche amplification drive transient current concentration, which further leads
to localized power dissipation and lattice-temperature rise \cite{Wang2024,WangNegGate2024}. 
The ion incident region, penetration depth, and device architecture further influence the spatial localization of electric field and thermal stress during burnout \cite{Germanicus2025,Yuan2024,Martinella2023}. 
Such information helps identify burnout-sensitive regions and guide device-level hardening.
However, these analyses mainly focus on failure mechanisms in specific device structures. 
The stochastic nature of burnout near the threshold has received less direct attention.

A related line of work has focused on radiation hardening and structural optimization of SiC power MOSFETs \cite{Wang2019,Bi2020,Sun2025,Kim2022,Liao2024}. 
These studies aim to reshape the electric field, reduce current concentration, and suppress
electrothermal runaway in specific device structures. 
However, these device-level and structure-oriented studies mainly address deterministic failure mechanisms under prescribed bias and strike conditions. 
The stochastic nature of burnout near the threshold has received less direct attention.
Probabilistic response metrics are widely used in single event effect testing \cite{Alberton2022,Sengupta2024}. 
These metrics describe ensemble level outcomes rather than the stochastic evolution of an individual electrothermal failure path.
Deterministic device simulations and empirical probability curves therefore remain only partially connected.

A compact stochastic dynamical formulation is therefore needed to connect the physical ingredients of burnout with boundary crossing and first passage observables. 
In a first-passage description, failure is treated as the arrival of a stochastic trajectory at an absorbing boundary rather than as a single deterministic switching event. 
This viewpoint is suitable for near-threshold burnout because it treats failure as a stochastic boundary-crossing event. 
It also provides a common framework for failure probability, survival statistics, and failure-time variability \cite{Redner2023, Redner2001}.

In this work, heavy-ion-induced burnout in SiC power MOSFETs is formulated as a stochastic first-passage problem in a reduced nonlinear
electrothermal system. 
The central contribution is to describe SEB as stochastic boundary crossing rather than as a purely deterministic
threshold event. 
The model treats ion-induced charge deposition and avalanche feedback as driving processes, while carrier loss and thermal relaxation provide the main relaxation channels within an absorbing-boundary framework.
This formulation links device-level burnout mechanisms to probabilistic failure observables, including survival behavior, first-passage timing,
and threshold broadening.
It also interprets noise-induced subthreshold burnout as a rare first-passage event. 
The resulting feedback--relaxation phase diagram separates recoverable, probabilistic, and rapidly unstable regimes.
The proposed framework provides a statistical-physics layer that complements TCAD simulations and irradiation testing by explaining the
observed burnout threshold as an emergent stochastic outcome, rather than as a fixed deterministic boundary.

The remainder of this paper is organized as follows.
Section~\ref{deterministic_model} develops the deterministic reduced electrothermal model and defines the relevant dimensionless
quantities. 
Section~\ref{stochastic_model} introduces the stochastic extension and formulates SEB as a first-passage problem with an absorbing
thermal boundary. 
Section~\ref{numerical_method} describes the Monte Carlo trajectory method and the statistical observables.
Section~\ref{results} presents the deterministic reference response, stochastic threshold broadening, subthreshold burnout, first-passage statistics, and the feedback--relaxation phase diagram.
Section~\ref{discussion} discusses the connection to SEB of SiC power MOSFETs, the distinction from deterministic TCAD threshold
analysis, and the limitations of the reduced model. Section~\ref{conclusion} summarizes the main findings.

\section{Deterministic reduced-order electrothermal model}
\label{deterministic_model}

SEB in a power device starts from localized ionization and develops through coupled electrothermal feedback
\cite{Grome2024, Wang2024, Mo2023,WangNegGate2024}. 
In a SiC power MOSFET, the relevant high-field region is typically located near the channel, junction field-effect transistor (JFET) region, P-well edge, or drift/buffer structure \cite{WangNegGate2024,Germanicus2025, Liao2024}.
The present reduced model does not explicitly resolve these geometric details.
Instead, their net effects are represented by a small set of effective parameters.
This reduction allows the burnout process to be formulated as a threshold-crossing problem for collective electrothermal variables. 
The deterministic response obtained from this model serves as a reference for the probabilistic burnout transition introduced by fluctuations.

\subsection{Sensitive-region approximation}
\label{subsensitive_region}

The model represents the spatially resolved device response by a small set of collective electrothermal variables. 
This reduction follows the projection-based viewpoint used in statistical physics \cite{Zwanzig1961,Mori1965}. 
The projection is performed over a localized active region, denoted by $\Omega_s$. 
In this region, ion-induced carrier generation can couple efficiently with electrothermal feedback.
Here, $\Omega_s$ denotes an effective active region rather than a resolved device geometry.
It is a coarse-grained sensitive volume used to define the effective degrees of freedom entering the reduced deterministic and stochastic dynamics~\cite{Petersen2011,Titus2013}.

\begin{figure}[t]
\centering
\includegraphics[width=0.6\textwidth]{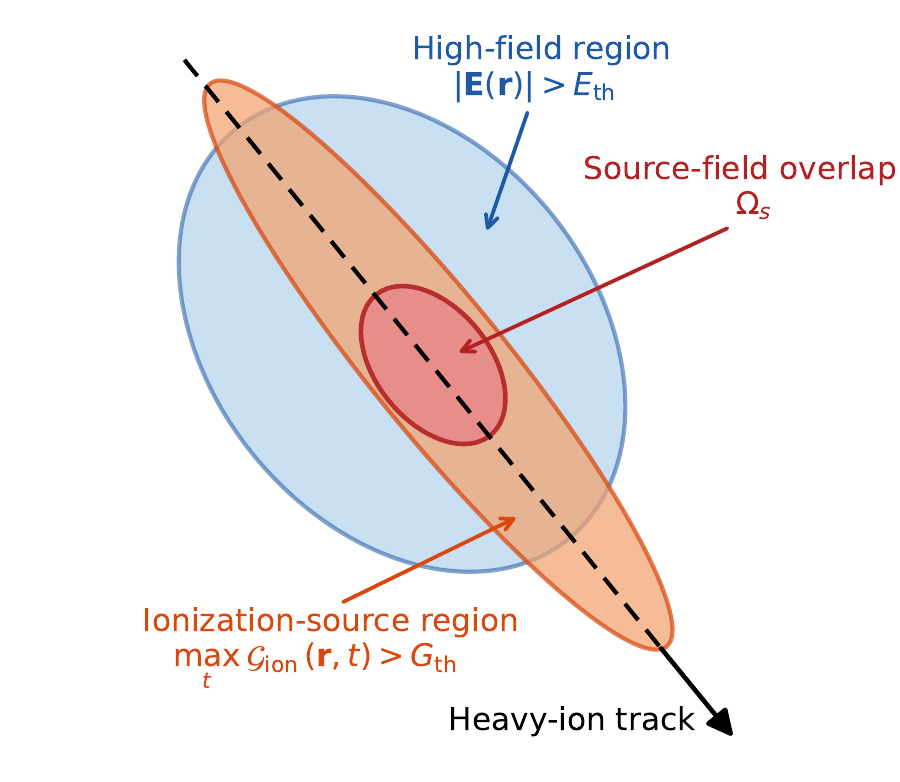}
\caption{Source and field overlap defining the coarse-grained sensitive region $\Omega_s$.}
\label{fig_sensitive_region}
\end{figure}

The sensitive region can be represented by the
sharp-region criterion
\begin{equation}
\Omega_s =\left\{\mathbf r: |\mathbf E(\mathbf r)|>E_{\rm th},
\quad
\max_t \mathcal{G}_{\rm ion}(\mathbf r,t)>G_{\rm th} \right\},
\label{Omega_definition}
\end{equation}
where the maximum is taken over the short ion-generation interval.
$E_{\rm th}$ gives the characteristic field scale for avalanche multiplication. 
The ionization source is normalized by the reference scale $G_{\rm th}$, while $\mathcal{G}_{\rm ion}(\mathbf r,t)$ represents
the source density generated by the ion track.
The region $\Omega_s$ therefore denotes the overlap between the high-field and ionization-source subregions. 
The corresponding projected source, $G_{\rm ion}(t)$, is defined in
Eq.~\eqref{Gion_projection}. 
Thermal confinement is represented by effective heat capacity and relaxation parameters. 
Figure~\ref{fig_sensitive_region} illustrates the overlap between the source region and the high-field region.

The threshold criteria define $\Omega_s$ for coarse graining, not as a sharp physical boundary.
More generally, the projection can be introduced through a dimensionless window function $w(\mathbf r)$, with $0\leq w(\mathbf r)\leq 1$. This function is localized near the overlap between the source region and the high-field region, as in spatial coarse-graining
descriptions~\cite{Hardy1982}.
The corresponding effective active volume is
\begin{equation}
V_s=\int w(\mathbf r) dV .
\label{Vs_window}
\end{equation}
The sharp-region approximation is recovered when $w(\mathbf r)=1$ for $\mathbf r\in\Omega_s$ and $w(\mathbf r)=0$ otherwise. 
In this limit, $V_s=\int_{\Omega_s}dV$, namely the geometrical volume of the threshold-defined sensitive region.

The effective excess carrier population in the active region is defined by
\begin{equation}
N(t)=\int w(\mathbf r)n_{\rm ex}(\mathbf r,t)\,dV ,
\label{N_coarse}
\end{equation}
where $n_{\rm ex}(\mathbf r,t)$ is the excess carrier density generated
by the ion strike and by subsequent field-assisted multiplication. The
coarse-grained temperature is taken as a heat-capacity-weighted average,
\begin{equation}
T(t)=\frac{\int w(\mathbf r)\rho_{\rm m}(\mathbf r)c_p(\mathbf r)T(\mathbf r,t)dV}{C_{\rm th}},
\label{T_coarse}
\end{equation}
with
\begin{equation}
C_{\rm th}=\int w(\mathbf r)\rho_{\rm m}(\mathbf r)c_p(\mathbf r)dV .
\label{Cth_general}
\end{equation}
Here $\rho_{\rm m}$ and $c_p$ denote the material mass density and specific heat capacity, respectively. 
For a homogeneous active region, Eq.~\eqref{Cth_general} reduces to
\begin{equation}
C_{\rm th}=\rho_{\rm m}c_p V_s .
\label{Cth_homogeneous}
\end{equation}
Thus, $N(t)$ and $T(t)$ are collective variables of the active region, not microscopic local fields.

The electric field entering the reduced model is represented by the coarse-grained magnitude
\begin{equation}
E_0 =\frac{\int w(\mathbf r)|\mathbf E(\mathbf r)|dV}{\int w(\mathbf r)dV}.
\label{E0_average}
\end{equation}
In an effective vertical blocking geometry, this field scale may be
approximated as
\begin{equation}
E_0 \simeq \beta_E\frac{V_{\rm DS}}{L_{\rm drift}},
\label{E0}
\end{equation}
where $V_{\rm DS}$ is the applied blocking voltage and $L_{\rm drift}$ is the characteristic blocking length used to estimate the average
blocking field~\cite{Baliga2008,Kimoto2014}.
The dimensionless factor $\beta_E$ accounts for local field enhancement caused by field crowding, junction curvature, and doping inhomogeneity. 
In the reduced model, $\beta_E$ is therefore treated as an effective feedback-control parameter rather than as a quantity obtained from a spatially resolved field solution.

Carrier removal is characterized by an effective loss time $\tau_{\rm loss}$, which represents carrier extraction,
recombination, and transport out of the active region. 
Heat removal is described separately by a lumped thermal resistance $R_{\rm th}$, giving the thermal relaxation time
\begin{equation}
\tau_{\rm th}=R_{\rm th}C_{\rm th}.
\label{thermal_time}
\end{equation}
Thus, $\tau_{\rm loss}$ and $\tau_{\rm th}$ represent the characteristic
carrier and thermal relaxation time scales of the reduced active-region
description.

This approximation is designed to describe threshold statistics rather than spatial profiles. 
Spatial inhomogeneity is absorbed into the effective parameters $V_s$, $\beta_E$, $\tau_{\rm loss}$, $C_{\rm th}$, and $R_{\rm th}$.
The localization of electrothermal work enters the reduced temperature dynamics through the heating term introduced later in
Section~\ref{subthermal_dynamics}, where the energy-localization coefficient $\zeta_E$ is defined. 
The irreversible failure condition is represented by an absorbing boundary in the coarse-grained temperature variable,
\begin{equation}
T(t)=T_{\rm c},
\label{absorbing_boundary}
\end{equation}
where $T_{\rm c}$ is an effective critical temperature associated with electrothermal damage.
Burnout is then formulated as a first-passage event in which the collective temperature $T(t)$ reaches $T_{\rm c}$, in the sense of
absorbing-boundary first-passage theory~\cite{Redner2001}. 
This construction provides the deterministic coarse-grained basis for the stochastic threshold-activation analysis developed in the following sections.

\subsection{Ionization source and carrier dynamics}
\label{subcarrier_dynamics}

The collective carrier variable $N(t)$ evolves under source injection, avalanche multiplication, and carrier removal from the active region.
The coarse-grained ionization source is obtained by projecting the spatially resolved source density onto the same active region used in
the sensitive-region approximation,
\begin{equation}
G_{\rm ion}(t)=\int w(\mathbf r)\mathcal{G}_{\rm ion}(\mathbf r,t)dV .
\label{Gion_projection}
\end{equation}
Thus, $G_{\rm ion}(t)$ is the total carrier-generation rate entering the
reduced carrier equation.

For a single ion strike, this source is represented by a finite-duration pulse,
\begin{equation}
G_{\rm ion}(t)=\frac{N_{\rm ion}}{\sqrt{2\pi}\tau_{\rm ion}}
\exp\left[-\frac{(t-t_s)^2}{2\tau_{\rm ion}^2}\right],
\label{Gion_dimensional}
\end{equation}
where $t_s$ is the strike time, $\tau_{\rm ion}$ is the pulse duration, and $N_{\rm ion}$ is the primary carrier population injected into the
coarse-grained active region. 
The normalization satisfies
\begin{equation}
\int_{-\infty}^{\infty}G_{\rm ion}(t)dt=N_{\rm ion}.
\label{Gion_normalization}
\end{equation}
For an observation window much longer than $\tau_{\rm ion}$ and centered on the strike, this normalization gives the total injected carrier
number to leading order.
In finite-time numerical calculations, the strike time is chosen sufficiently far from the initial boundary, or the pulse is normalized over the computational time window.

The primary carrier population is estimated from the ionization energy deposited within the active region~\cite{Petersen2011,Titus2013, Messenger1992}. 
If the linear energy transfer (LET) is given as a mass stopping power, the deposited energy per unit path length is given by
${\rm LET}\rho_{\rm m}$, where $\rho_{\rm m}$ is the mass density of the material. 
A coarse-grained estimate is
\begin{equation}
N_{\rm ion}=\xi_{\rm dep}\frac{{\rm LET}\,\rho_{\rm m}\,L_{\rm eff}(\theta)}{\varepsilon_{\rm eh}},
\label{Nion_physical}
\end{equation}
where $\varepsilon_{\rm eh}$ is the mean energy required to create one electron-hole pair and $\xi_{\rm dep}$ is a dimensionless deposition
and overlap factor. 
This factor accounts for finite-range and spreading effects that reduce the coupling between the ion track and the active region.
Equivalently,
\begin{equation}
N_{\rm ion}=\chi_{\rm ion}{\rm LET}L_{\rm eff}(\theta),
\label{Nion}
\end{equation}
where $\chi_{\rm ion}$ collects the material and deposition factors, like pair-creation energy and deposition efficiency, needed to convert LET into an effective carrier source.
The effective track length is modeled as
\begin{equation}
L_{\rm eff}(\theta)=\min\left(\frac{L_s}{|\cos\theta|},L_{\max}\right),
\label{Leff}
\end{equation}
where $L_s$ is the normal length scale of the active region and $L_{\max}$ represents the finite device thickness, finite ion range, or
loss of overlap with the high-field domain. 
This expression is used as a coarse-grained geometrical factor rather than as a detailed ion-transport model.

The deterministic carrier dynamics are written as
\begin{equation}
\frac{dN}{dt}=G_{\rm ion}(t)+\Gamma[E(t),T(t)]N(t)-\frac{N(t)}{\tau_{\rm loss}}.
\label{N_dimensional}
\end{equation}
The three terms represent primary carrier injection, avalanche multiplication, and effective carrier loss, respectively. 
The loss term uses the same $\tau_{\rm loss}$ introduced in the active-region description.

The avalanche multiplication rate is modeled as
\begin{equation}
\Gamma[E(t),T(t)]=v_s\alpha[E(t),T(t)],
\label{Gamma_ava}
\end{equation}
where $v_s$ is an effective saturated carrier velocity and $\alpha(E,T)$ is an impact-ionization coefficient~\cite{Baliga2008,Kimoto2014,Sze2006}. 
To retain the strong field sensitivity of avalanche multiplication in a reduced form, we use
\begin{equation}
\alpha(E,T)=\alpha_0\exp\left[-\frac{E_c(T)}{E}\right],
\label{alpha}
\end{equation}
with
\begin{equation}
E_c(T)=E_{c0}\left[1+a_T(T-T_0)\right].
\label{EcT}
\end{equation}
Here $\alpha_0$ is a reference impact-ionization prefactor, $E_{c0}$ is
a reference characteristic field, and $a_T$ parameterizes the thermal
dependence of the effective avalanche coefficient. 
These quantities are treated as effective parameters of the reduced model rather than
universal material constants. 
For $a_T>0$, increasing temperature raises the effective characteristic field and weakens the impact-ionization
coefficient. 
The electrothermal feedback considered here arises primarily from carrier multiplication, field redistribution, and localized Joule heating.

The field entering the feedback term is linked to the coarse-grained field scale $E_0$. To represent carrier-induced field redistribution in
a bounded phenomenological form, we write
\begin{equation}
E(t)=E_0\left[1+\eta_n\frac{N(t)}{N(t)+N_s}\right],
\label{Efield_dynamic}
\end{equation}
where $\eta_n$ is a dimensionless feedback parameter and $N_s$ is a saturation carrier number. 
Positive $\eta_n$ represents local field enhancement, whereas negative $\eta_n$ represents field screening. 
The limit $\eta_n=0$ gives a fixed-field approximation, in which avalanche multiplication remains controlled by $E_0$ and carrier-induced field redistribution is neglected. 
These deterministic relations close the carrier sector of the reduced electrothermal model and provide the drift part of the stochastic carrier dynamics introduced below.

\subsection{Thermal dynamics and deterministic burnout indicators}
\label{subthermal_dynamics}

The coarse-grained temperature $T(t)$ evolves according to an energy-balance equation for the active region. 
The deterministic temperature balance is written as
\begin{equation}
C_{\rm th}\frac{dT}{dt}=P(t)-\frac{T(t)-T_0}{R_{\rm th}},
\label{T_dimensional}
\end{equation}
where $T_0$ is the ambient or substrate temperature. 
The first term on the right-hand side represents localized electrothermal heating, and
the second term gives heat removal through the effective thermal resistance $R_{\rm th}$.

At the spatial level, the electrothermal heating power in the active region is represented by the weighted Joule source~\cite{Baliga2008, Sze2006}
\begin{equation}
P_{\Omega}(t)=\int w(\mathbf r)\mathbf J(\mathbf r,t)\cdot \mathbf E_{\rm loc}(\mathbf r,t)dV .
\label{power_integral}
\end{equation}
Here $\mathbf J(\mathbf r,t)$ denotes the transient current density associated with the ion-induced and avalanche-amplified carrier
population, while $\mathbf E_{\rm loc}(\mathbf r,t)$ is the corresponding local electric field used to evaluate the spatial Joule source. 

In the reduced description, $\mathbf E_{\rm loc}(\mathbf r,t)$ is replaced by an effective heating power expressed in terms of the same
scalar coarse-grained field magnitude $E(t)$ used in the carrier dynamics,
\begin{equation}
P(t)=q\zeta_E\,v_s\,E(t)N(t),
\label{power_dimensional}
\end{equation}
where $q$ is the elementary charge. 
The coefficient $\zeta_E$ is an energy-localization factor that accounts for current confinement and carrier--field coupling in local heat generation.
Larger $\zeta_E$ indicates stronger localized heat production, whereas smaller $\zeta_E$ represents current spreading. 
It is distinct from $\eta_n$, which represents carrier-driven changes in the coarse-grained electric field.  
This separation keeps field feedback and heat localization as independent mechanisms in the reduced model.

Within the constant effective heat-capacity approximation, the thermal
energy margin associated with the absorbing boundary is
\begin{equation}
W_{\rm th}=C_{\rm th}(T_{\rm c}-T_0),
\label{Wth}
\end{equation}
and the electrothermal work deposited during the observation window
$t_c$ is
\begin{equation}
W_{\rm dep}(t_c)=\int_0^{t_c}P(t)\,dt .
\label{Wdep}
\end{equation}
The deposited-work factor is then defined as
\begin{equation}
\Lambda_{\rm dep}=\frac{W_{\rm dep}(t_c)}{W_{\rm th}}.
\label{Lambda_dep}
\end{equation}
This quantity measures the accumulated electrothermal input relative to
the thermal margin. Since heat is simultaneously removed through the
relaxation term in Eq.~\eqref{T_dimensional}, $\Lambda_{\rm dep}$
does not by itself define the burnout boundary.

For the initial condition $T(0)=T_0$, Eq.~\eqref{T_dimensional} gives
\begin{equation}
T(t)-T_0=\frac{1}{C_{\rm th}}\int_0^t\exp\left[-\frac{t-u}{R_{\rm th}C_{\rm th}}\right]P(u)du .
\label{T_convolution_dimensional}
\end{equation}
The exponential kernel expresses thermal relaxation of earlier heat deposition. 
The peak normalized temperature is then measured by
\begin{equation}
\Lambda_{\rm th}=\max_{0\le t\le t_c}\frac{T(t)-T_0}{T_{\rm c}-T_0}.
\label{Lambda_th_dimensional}
\end{equation}
The deterministic burnout condition is
\begin{equation}
\Lambda_{\rm th}\ge 1.
\label{deterministic_boundary}
\end{equation}
Values below unity correspond to subcritical trajectories, while values at or above unity indicate that the absorbing boundary has been reached.

The two indicators $\Lambda_{\rm dep}$ and $\Lambda_{\rm th}$ describe different aspects of the deterministic thermal response. 
The deposited-work factor $\Lambda_{\rm dep}$ summarizes the accumulated electrothermal input, while
$\Lambda_{\rm th}$ includes the competition between heat deposition and thermal relaxation. 
In the stochastic formulation below, burnout is imposed through the first-passage condition $T(t)=T_{\rm c}$, and
$\Lambda_{\rm dep}$ is used only as a coarse-grained energy-input indicator~\cite{Redner2001}.

\subsection{Dimensionless form}
\label{subdimensionless_form}

A dimensionless representation separates the stochastic-threshold structure from device-dependent scales. Time,
carrier population, and temperature are rescaled as
\begin{equation}
s=\frac{t}{\tau_{\rm loss}},
\qquad
n(s)=\frac{N(t)}{N_*},
\qquad
\Theta(s)=\frac{T(t)-T_0}{T_{\rm c}-T_0},
\label{dimensionless_variables}
\end{equation}
where $N_*$ is a reference carrier number. The absorbing boundary
$T=T_{\rm c}$ is then written as
\begin{equation}
\Theta=1.
\label{dimensionless_absorbing_boundary}
\end{equation}

The dimensionless ionization source is defined by $g(s;\ell)=\tau_{\rm loss}G_{\rm ion}(t)/N_*$. 
Substitution of Eq.~\eqref{Gion_dimensional} gives
\begin{equation}
g(s;\ell)=\frac{\ell}{\sqrt{2\pi}\sigma_s}\exp\left[-\frac{(s-s_0)^2}{2\sigma_s^2}\right],
\label{g_dimensionless}
\end{equation}
with
\begin{equation}
s_0=\frac{t_s}{\tau_{\rm loss}},
\qquad
\sigma_s=\frac{\tau_{\rm ion}}{\tau_{\rm loss}},
\qquad
\ell=\frac{N_{\rm ion}}{N_*}.
\label{ell_definition}
\end{equation}
The parameter $\ell$ is therefore the dimensionless primary ionization strength and is proportional to ${\rm LET}\,L_{\rm eff}(\theta)$
through Eq.~\eqref{Nion}.

The scalar field entering the carrier and thermal equations is rescaled as
\begin{equation}
e(s)=\frac{E(t)}{E_{\rm ref}},
\label{e_definition}
\end{equation}
where $E_{\rm ref}$ is a reference field scale. Using the bounded field-redistribution closure, the dimensionless field is
\begin{equation}
e(s)=b\left[1+\eta_n\frac{n(s)}{n(s)+n_s}\right],
\label{e_dimensionless}
\end{equation}
where
\begin{equation}
b=\frac{E_0}{E_{\rm ref}}
\simeq
\frac{\beta_E V_{\rm DS}}{E_{\rm ref}L_{\rm drift}},
\qquad
n_s=\frac{N_s}{N_*}.
\label{b_ns_definition}
\end{equation}
Here $b$ sets the externally imposed electric field, while $\eta_n$ determines the field change produced by excess carriers.

The dimensionless avalanche feedback is
\begin{equation}
f(e,\Theta)=\tau_{\rm loss}\Gamma[E(t),T(t)].
\label{f_dimensionless}
\end{equation}
Using $E(t)=E_{\rm ref}e(s)$ and $T(t)=T_0+(T_{\rm c}-T_0)\Theta(s)$, the reduced Chynoweth-type coefficient can be written as
\begin{equation}
f(e,\Theta)=A_f\exp\left[-\frac{B_f(1+a_{\Theta}\Theta)}{e}\right],
\label{f_explicit}
\end{equation}
where
\begin{equation}
A_f=\tau_{\rm loss}v_s\alpha_0,
\qquad
B_f=\frac{E_{c0}}{E_{\rm ref}},
\qquad
a_{\Theta}=a_T(T_{\rm c}-T_0).
\label{f_parameters}
\end{equation}
This form retains the strong field sensitivity of avalanche multiplication while absorbing material and device scales into a small
number of dimensionless parameters.

The carrier equation becomes
\begin{equation}
\frac{dn}{ds}=g(s;\ell)+\left[f(e,\Theta)-1\right]n.
\label{n_dimensionless}
\end{equation}
The term $f(e,\Theta)n$ represents avalanche amplification and the term $-n$ is carrier removal on the loss time scale
$\tau_{\rm loss}$.

The temperature equation takes the dimensionless form
\begin{equation}
\frac{d\Theta}{ds}=\kappa e(s)n(s)-r\Theta(s),
\label{theta_dimensionless}
\end{equation}
where
\begin{equation}
\kappa=\frac{q\zeta_E v_s E_{\rm ref}N_*\tau_{\rm loss}}{C_{\rm th}(T_{\rm c}-T_0)},
\qquad
r=\frac{\tau_{\rm loss}}{R_{\rm th}C_{\rm th}}.
\label{kappa_r}
\end{equation}
The parameter $\kappa$ measures the electrothermal energy injected over one carrier-loss time relative to the thermal margin. 
The energy-localization coefficient $\zeta_E$ enters the dimensionless
dynamics through $\kappa$, whereas $\eta_n$ enters through the feedback field $e(s)$. 
The parameter $r$ is the thermal relaxation rate measured relative to the carrier-loss rate.

The deposited-work indicator becomes
\begin{equation}
\Lambda_{\rm dep}=\int_0^{s_c}\kappa e(s)n(s)ds,
\label{lambda_dep_dimensionless}
\end{equation}
where $s_c=t_c/\tau_{\rm loss}$. 
The thermal indicator associated with the absorbing boundary is
\begin{equation}
\Lambda_{\rm th}=\max_{0\le s\le s_c}\Theta(s).
\label{lambda_th_dimensionless}
\end{equation}
For $\Theta(0)=0$, Eq.~\eqref{theta_dimensionless} gives
\begin{equation}
\Theta(s)=\int_0^s\exp[-r(s-u)]\kappa e(u)n(u)du .
\label{theta_convolution_dimensionless}
\end{equation}
The deterministic failure condition is $\Lambda_{\rm th}\ge 1$.

\begin{figure}[t]
\centering
\includegraphics[width=0.95\textwidth]{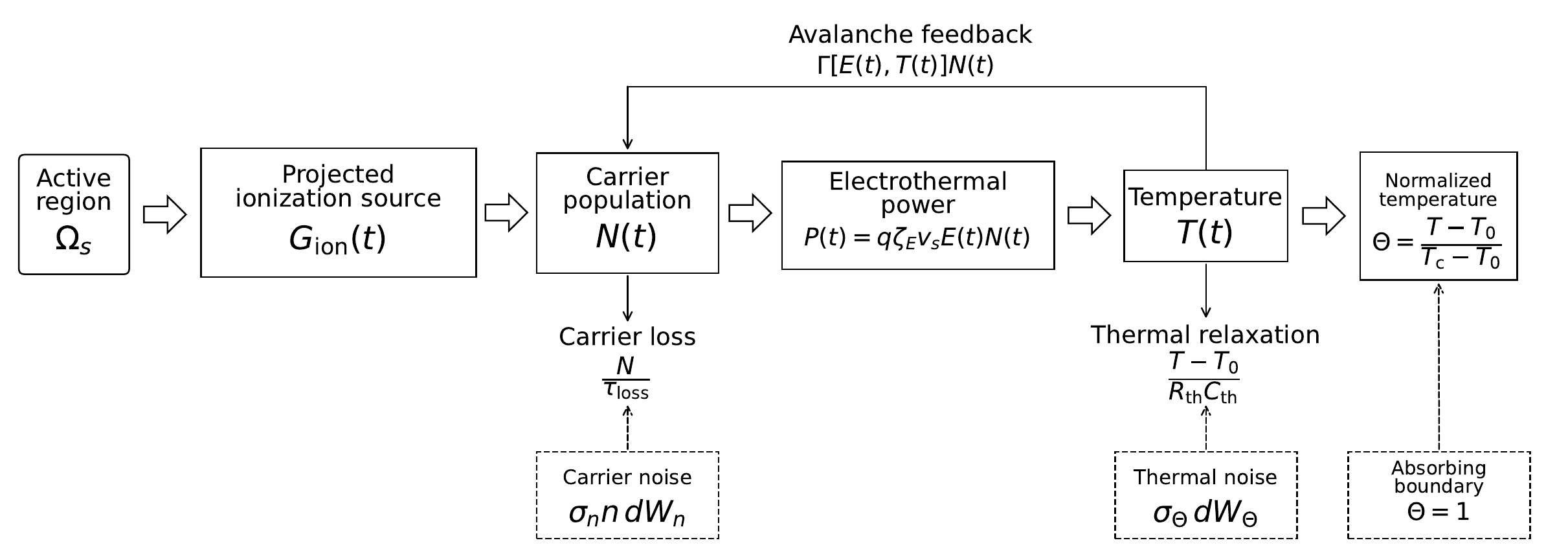}
\caption{Reduced electrothermal framework. The upper solid-line part represents the deterministic reduced-order model, while the lower dashed-line part represents the stochastic extension.}
\label{fig_reduced_framework}
\end{figure}

The dimensionless system is controlled by three groups of parameters.
The ionization strength $\ell$ and the imposed field parameter $b$ specify the external excitation. 
The avalanche function $f(e,\Theta)$, the energy injection scale $\kappa$, and the coefficient $\eta_n$ describe electrothermal feedback. The thermal relaxation rate $r$ sets the main recovery channel.
These variables define the parameter space used in Sections~\ref{substochastic_broadening} and \ref{subphase_diagram} to construct the stochastic probability curves and the feedback-relaxation phase diagram.
Figure~\ref{fig_reduced_framework} summarizes the reduced
electrothermal construction. 
The deterministic sector links ionization, carrier multiplication, localized heating, and thermal relaxation to the
normalized temperature variable. 
The stochastic extension in the next section introduces fluctuations, with burnout formulated as first-passage to an absorbing boundary.

\section{Stochastic threshold-activation model}
\label{stochastic_model}

Radiation-induced electrothermal failure is sensitive to event-level variability. 
Fluctuations in energy deposition, ion-track geometry, and local device conditions can alter the failure trajectory.
In the reduced description, these unresolved degrees of freedom are represented by stochastic terms added to the dimensionless
carrier and temperature equations. 
The It\^{o} stochastic dynamics are written as~\cite{Gardiner2009,VanKampen2007}
\begin{equation}
dn=\left[g(s;\ell)+\left(f(e,\Theta)-1\right)n\right]ds+\sigma_n n dW_n(s),
\label{sde_n}
\end{equation}
and
\begin{equation}
d\Theta=\left[\kappa e(s)n-r\Theta\right]ds+\sigma_\Theta dW_\Theta(s).
\label{sde_theta}
\end{equation}
Here $W_n(s)$ and $W_\Theta(s)$ are Wiener processes in the dimensionless time $s$. 
The coefficient $\sigma_n$ controls multiplicative fluctuations in the carrier population, while $\sigma_\Theta$ sets the strength of
unresolved thermal variability.
The factor $n$ in $\sigma_n n dW_n$ reflects the assumption that carrier fluctuations scale with the instantaneous excess carrier population. The field $e(s)$ is evaluated from the state-dependent feedback relation in Eq.~\eqref{e_dimensionless}.

Unless stated otherwise, $W_n(s)$ and $W_\Theta(s)$ are independent
standard Wiener processes satisfying~\cite{Gardiner2009,VanKampen2007}
\begin{equation}
\mathbb E[dW_i(s)] = 0, \qquad
\mathbb E[dW_i(s)dW_j(s)] = \delta_{ij}\,ds,
\qquad i,j\in\{n,\Theta\}.
\label{wiener_basic}
\end{equation}
A correlated extension can be introduced by replacing the off-diagonal
covariance with
\begin{equation}
\mathbb E[dW_n(s)dW_\Theta(s)] = \rho_{n\Theta}ds,
\qquad -1\le \rho_{n\Theta}\le 1,
\label{correlated_noise}
\end{equation}
while retaining
\begin{equation}
\mathbb E[dW_n^2(s)]=\mathbb E[dW_\Theta^2(s)]=ds.
\end{equation}
The baseline model uses the independent case to separate carrier and thermal noise effects without introducing an additional correlation
parameter.

The drift terms are
\begin{equation}
A_n(n,\Theta,s)=g(s;\ell)+\left[f(e,\Theta)-1\right]n,
\label{An_definition}
\end{equation}
and
\begin{equation}
A_\Theta(n,\Theta,s)=\kappa e(s)n-r\Theta .
\label{Atheta_definition}
\end{equation}
In these drift terms, $e(s)$ is understood as the feedback field determined by the current carrier state. 
For the independent-noise baseline, the probability density $p(n,\Theta,s)$ satisfies the corresponding Fokker-Planck equation
\cite{Gardiner2009, Risken1996}
\begin{align}
\frac{\partial p}{\partial s}=&
-\frac{\partial}{\partial n}\left[A_n(n,\Theta,s)p\right]-\frac{\partial}{\partial \Theta}\left[A_\Theta(n,\Theta,s)p\right]\nonumber\\
&+\frac{1}{2}\frac{\partial^2}{\partial n^2}\left[\sigma_n^2 n^2 p\right]+\frac{1}{2}\sigma_\Theta^2
\frac{\partial^2 p}{\partial \Theta^2}.
\label{fokker_planck}
\end{align}
The numerical calculations in this work use independent noise with $\rho_{n\Theta}=0$. 
This choice isolates the separate effects of carrier and thermal fluctuations without introducing an uncalibrated correlation parameter. 
A correlated extension would require an additional mixed diffusion term and is not considered below. 
This Fokker-Planck form identifies the probability currents and the state-space domain associated with the stochastic reduced model.

The probability domain is
\begin{equation}
n\ge 0,
\qquad
\Theta_{\min}\le \Theta <1,
\label{probability_domain}
\end{equation}
with
\begin{equation}
\Theta_{\min}=-\frac{T_0}{T_{\rm c}-T_0}.
\label{theta_min_stochastic}
\end{equation}
The lower bound corresponds to $T=0$. In the parameter regime considered here, $\sigma_\Theta$ is chosen so that this lower boundary is not dynamically active. 
It is retained only to make the physical temperature domain explicit.

For a single-pulse reference state, the stochastic trajectories start from
\begin{equation}
n(0)=0,
\qquad
\Theta(0)=0.
\label{sde_initial_condition}
\end{equation}
The corresponding Fokker--Planck initial condition is
\begin{equation}
p(n,\Theta,0)=\delta(n)\delta(\Theta).
\label{fp_initial_condition}
\end{equation}
Because the carrier noise is multiplicative, it vanishes at $n=0$ and does not generate carriers before the ionization pulse. 
Carrier injection occurs through the source term $g(s;\ell)$, after which the multiplicative noise modulates avalanche growth and carrier removal.

At the thermal failure boundary, the absorbing condition is
\begin{equation}
p(n,\Theta=1,s)=0.
\label{fp_absorbing_boundary}
\end{equation}
This boundary represents irreversible burnout in the coarse-grained temperature variable.
At $n=0$, the carrier population is constrained to remain at or above zero. 
After the ionization source has vanished, the continuous carrier dynamics may be expressed by the zero flux condition
\begin{equation}
J_n(n=0,\Theta,s)=0,
\qquad
g(s;\ell)\simeq 0,
\label{n_no_flux_limited}
\end{equation}
where
\begin{equation}
J_n=A_n p-\frac{1}{2}\frac{\partial}{\partial n}\left(\sigma_n^2 n^2 p \right)
\label{Jn_current}
\end{equation}
is the carrier probability current. During the pulse, the source term moves probability into the positive-$n$ domain. 
In the trajectory implementation, the carrier population is constrained to remain at or above zero after each Euler-Maruyama update by the
projection step in Eq.~\eqref{n_projection}.

The lower temperature boundary $\Theta=\Theta_{\min}$ is not an absorbing failure boundary. 
It may be treated as a reflecting physical bound if thermal fluctuations are large enough to reach it. 
In the numerical calculations considered here, crossings below $\Theta_{\min}$ are statistically negligible, and the lower projection
serves only as a numerical safeguard.

The first-passage time to burnout is defined by~\cite{Redner2001}
\begin{equation}
\tau_{\rm FPT}=\inf\{s>0:\Theta(s)\ge 1\}.
\label{fpt}
\end{equation}
If the trajectory remains below the absorbing boundary during
$0\le s\le s_c$, the first-passage time is censored at $s_c$. The
burnout probability over the observation window is
\begin{equation}
P_{\rm SEB}=\mathbb P\left[\tau_{\rm FPT}\le s_c\right].
\label{pseb_short}
\end{equation}
Equivalently,
\begin{equation}
P_{\rm SEB}=\mathbb P\left[\max_{0\le s\le s_c}\Theta(s)\ge 1\right].
\label{pseb}
\end{equation}

For compact notation, the fixed pulse-shape and field-redistribution
parameters are grouped as
\begin{equation}
\mathcal P=(\eta_n,n_s,s_0,\sigma_s,s_c).
\label{fixed_parameter_set}
\end{equation}
The burnout probability is written as
\begin{equation}
P_{\rm SEB}=P_{\rm SEB}\left(\ell,b,\kappa,r,\sigma_n,\sigma_\Theta,\mathcal P \right).
\label{pseb_parameter_compact}
\end{equation}
Here $\ell$ and $b$ specify the external excitation through ionization
strength and imposed field scale. 
The parameter $\kappa$ measures electrothermal energy injection, while $r$ gives the thermal relaxation rate. 
The noise amplitudes are set by $\sigma_n$ for carrier fluctuations and $\sigma_\Theta$ for thermal
fluctuations.
$\mathcal P$ is held fixed unless otherwise stated. 
The energy-localization coefficient $\zeta_E$ does not appear as an additional independent argument because
it is absorbed into $\kappa$ through Eq.~\eqref{kappa_r}. 
Varying $\zeta_E$ in physical parameters therefore corresponds to varying $\kappa$ in the dimensionless model.

In the zero-noise limit, the stochastic formulation reduces to the deterministic boundary $\Lambda_{\rm th}=1$. 
Finite $\sigma_n$ or $\sigma_\Theta$ broadens this deterministic boundary into a probabilistic transition band. 
At fixed $b$, $\kappa$, $r$, $\sigma_n$, $\sigma_\Theta$, and $\mathcal P$, the transition width in ionization
strength is measured by
\begin{equation}
\Delta \ell=\ell_{0.9}-\ell_{0.1}.
\label{transition_width}
\end{equation}
The quantiles are defined through
\begin{equation}
P_{\rm SEB}(\ell_q)=q,
\qquad q\in\{0.1,0.9\},
\label{ell_quantiles}
\end{equation}
Thus, $\Delta\ell$ provides a compact measure of stochastic threshold broadening.

The survival probability is
\begin{equation}
S(s)=\mathbb P(\tau_{\rm FPT}>s),
\label{survival_probability}
\end{equation}
and the conditional mean first-passage time for failed trajectories is
\begin{equation}
\langle \tau_{\rm FPT}\rangle_{\rm fail}=\mathbb E[\tau_{\rm FPT}\mid \tau_{\rm FPT}\le s_c].
\label{conditional_mean_fpt}
\end{equation}
These observables separate rapid runaway from delayed noise-assisted failure and provide the basis for the numerical analysis. 
In the numerical implementation, $\ell_{0.1}$ and $\ell_{0.9}$ are obtained by interpolating the estimated probability curve $P_{\rm SEB}(\ell)$. 
This interpolation is used only to extract $\Delta\ell$ for comparing transition sharpness under different noise and relaxation conditions.

\section{Numerical method and observables}
\label{numerical_method}

The stochastic differential Eqs. \eqref{sde_n} and \eqref{sde_theta} are evaluated by ensemble simulation. 
Since the noise terms are written in It\^{o} form, the baseline integration uses the Euler-Maruyama scheme. Let $s_k=k\Delta s$, with
$k=0,1,\ldots,k_c$ and $s_c=k_c\Delta s$. 
The discrete variables are $n_k=n(s_k)$ and $\Theta_k=\Theta(s_k)$. 
At each time step, the feedback field and avalanche factor are evaluated as
\begin{equation}
e_k=b\left[1+\eta_n\frac{n_k}{n_k+n_s}\right],
\label{e_k}
\end{equation}
and
\begin{equation}
f_k=f(e_k,\Theta_k).
\label{f_k}
\end{equation}
The Euler-Maruyama updates are
\begin{equation}
n_{k+1}=n_k+\left[g(s_k;\ell)+(f_k-1)n_k\right]\Delta s+\sigma_n n_k\sqrt{\Delta s}\xi_{n,k},
\label{EM_n}
\end{equation}
and
\begin{equation}
\Theta_{k+1}=\Theta_k+\left[\kappa e_k n_k-r\Theta_k\right]\Delta s+\sigma_\Theta\sqrt{\Delta s}\xi_{\Theta,k}.
\label{EM_theta}
\end{equation}
Here $\xi_{n,k}$ and $\xi_{\Theta,k}$ are independent standard normal random variables, so that $\sqrt{\Delta s}\,\xi_{n,k}$ and
$\sqrt{\Delta s}\,\xi_{\Theta,k}$ represent the discrete Wiener increments. 
For correlated carrier-thermal fluctuations, the two random variables may be sampled from a bivariate normal distribution
with correlation coefficient $\rho_{n\Theta}$. 
The independent case is used as the reference case.

The initial state is
\begin{equation}
n_0=0,
\qquad
\Theta_0=0.
\label{initial_conditions}
\end{equation}
This state has no excess carriers generated by the ion, and its coarse-grained temperature equals the ambient value before the strike. 
The multiplicative carrier noise vanishes at $n=0$, so stochastic carrier fluctuations do not create carriers before the ionization pulse. 
After source injection, the same noise term modulates avalanche growth and carrier removal.

The explicit update in Eq.~\eqref{EM_n} can generate small negative values of $n_k$ at finite time step. 
Since $n_k$ represents a carrier population, the numerical scheme enforces
\begin{equation}
n_{k+1}\leftarrow \max(n_{k+1},0).
\label{n_projection}
\end{equation}
This projection is a numerical regularization of the explicit scheme, not an additional physical drift. 
Its influence is checked by reducing $\Delta s$ and confirming that probability curves and first-passage statistics remain unchanged within sampling uncertainty.

For the thermal variable, the baseline noise amplitude is chosen so that crossings below $\Theta_{\min}$ are statistically negligible. 
If larger thermal-noise amplitudes are tested, the lower physical bound can be imposed by
\begin{equation}
\Theta_{k+1}\leftarrow \max(\Theta_{k+1},\Theta_{\min}).
\label{theta_projection}
\end{equation}
This lower projection is used only as a numerical safeguard.

The absorbing boundary is imposed through
\begin{equation}
\Theta_k\ge 1.
\label{discrete_absorbing}
\end{equation}
A trajectory is counted as a burnout event if this condition is reached for some $k\le k_c$. 
The discrete first-passage time of trajectory $i$ can be written as
\begin{equation}
\tau_{{\rm FPT}}^{(i)}=\min\{s_k:\Theta_k^{(i)}\ge 1\}.
\label{discrete_fpt}
\end{equation}
Trajectories that remain below the boundary until $s_c$ are censored at $s_c$.

For an ensemble of $M$ trajectories, the burnout indicator is given as
\begin{equation}
I_i=
\begin{cases}
1, & \tau_{{\rm FPT}}^{(i)}\le s_c,\\
0, & \tau_{{\rm FPT}}^{(i)}>s_c.
\end{cases}
\label{indicator}
\end{equation}
The Monte Carlo (MC) estimator of the burnout probability is
\begin{equation}
\widehat P_{\rm SEB} =\frac{1}{M}\sum_{i=1}^{M} I_i.
\label{p_estimator}
\end{equation}
The binomial standard error is
\begin{equation}
{\rm SE}\!\left(\widehat P_{\rm SEB}\right)=\sqrt{\frac{\widehat P_{\rm SEB}\left(1-\widehat P_{\rm SEB}\right)}{M}}.
\label{p_standard_error}
\end{equation}
The number of failed trajectories is
\begin{equation}
M_{\rm fail}=\sum_{i=1}^M I_i.
\label{mfail}
\end{equation}
The conditional mean first-passage time is evaluated as
\begin{equation}
\left\langle \tau_{\rm FPT}\right\rangle_{\rm fail}=\frac{1}{M_{\rm fail}}\sum_{i:I_i=1}\tau_{{\rm FPT}}^{(i)},
\qquad
M_{\rm fail}>0 .
\label{mean_fpt}
\end{equation}
The empirical survival probability is
\begin{equation}
\widehat S(s)=\frac{1}{M}\sum_{i=1}^{M}\mathbf 1\left(\tau_{{\rm FPT}}^{(i)}>s\right),
\label{empirical_survival}
\end{equation}
where $\mathbf 1(\cdot)$ denotes the indicator function. 
The first-passage-time distribution is obtained from the crossing times of the failed trajectories.

All numerical values are normalized model parameters rather than fitted parameters for a specific commercial SiC power MOSFET. 
They are selected to place the reduced electrothermal system near the feedback--relaxation transition. 
This regime allows the model to capture the change from recovery to probabilistic burnout and rapid runaway.
Such a parameter choice is consistent with the aim of the present work. 
Here the focus is on the stochastic threshold structure of the reduced model, rather than on a device-specific burnout voltage.
Quantitative prediction for a specific device would require calibration of the effective parameters against TCAD simulations or irradiation data.
The connection to SiC power-device SEB and the distinction from deterministic TCAD threshold analysis are discussed further in
Sections~\ref{sic_mapping} and~\ref{tcad_difference}, respectively.

Unless otherwise stated, the numerical results in Section~\ref{results} use the baseline normalized parameter set
listed in Table~\ref{numerical_baseline_parameters}. 
These values are not fitted device parameters. 
They are selected to place the reduced system near the feedback-relaxation transition and to keep the
dimensionless dynamics in a numerically resolved regime. 
The ionization pulse is centered at $s_0=2$ with width $\sigma_s=0.18$, making the
excitation short compared with the observation window $s_c=12$. 
The baseline field parameter is $b=1.20$, while $\eta_n=0.35$ and $n_s=1$ give moderate carrier driven field variation without a singular response at small carrier density. 
The thermal parameters are $\kappa=0.16$ and $r=0.18$, placing energy injection and thermal relaxation on comparable
normalized time scales. 
This choice allows both recovery and boundary crossing to occur within the simulated window. 
The avalanche feedback parameters are $A_f=5.50$, $B_f=2.20$, and $a_\Theta=0.12$, which set the amplification scale, field sensitivity, and temperature dependence of the effective ionization field. 
The use of $\ell$, $b$, $\kappa$, and $r$ as control parameters is motivated by heavy-ion and TCAD studies of SiC power MOSFETs, where SEB is governed by ionization strength, strike geometry, field enhancement, impact ionization, local heating, and thermal recovery
\cite{Wang2024,WangNegGate2024,Germanicus2025, Liao2024}.

\begin{table}[t]
\centering
\caption{Baseline normalized parameters used in the numerical
calculations.}
\label{numerical_baseline_parameters}
\begin{tabular}{lll}
\toprule
Parameter & Value & Role in the reduced model \\
\midrule
$s_0$ & $2$ & Center of the normalized ionization pulse \\
$\sigma_s$ & $0.18$ & Width of the normalized ionization pulse \\
$b$ & $1.20$ & Baseline field-control parameter \\
$\eta_n$ & $0.35$ & Carrier-induced field redistribution strength \\
$n_s$ & $1$ & Saturation scale in the field-redistribution term \\
$\kappa$ & $0.16$ & Electrothermal energy-injection scale \\
$r$ & $0.18$ & Thermal relaxation strength \\
$A_f$ & $5.50$ & Avalanche-amplification scale \\
$B_f$ & $2.20$ & Field sensitivity of avalanche feedback \\
$a_\Theta$ & $0.12$ & Temperature dependence of the feedback barrier \\
$s_c$ & $12$ & Observation window for first-passage events \\
\bottomrule
\end{tabular}
\end{table}

For stochastic simulations in Section~\ref{results}, the baseline time step is $\Delta s=0.006$, and the ensemble size is $M=3500$. 
The same numerical setting is used for all probability, first-passage, and phase-map calculations.
The convergence of these numerical choices is examined in Section~\ref{subnumerical_convergence}.
For visualization, the noise amplitudes are linked through
\begin{equation}
\sigma_n=\sigma,
\qquad
\sigma_\Theta=0.35\sigma .
\label{numerical_noise_convention}
\end{equation}
This convention makes carrier fluctuations the dominant stochastic contribution while retaining a smaller thermal fluctuation. 
The factor $0.35$ is a normalized modeling choice, not a fitted device parameter.
For the baseline ensemble size, the maximum binomial standard error of $\widehat P_{\rm SEB}$ occurs near $\widehat P_{\rm SEB}=0.5$ and is approximately
\begin{equation}
{\rm SE}_{\max}=\frac{1}{2\sqrt{M}}\simeq 0.0085.
\label{max_binomial_error}
\end{equation}
This uncertainty is small compared with the probability variations in the transition curves and phase maps reported in
Sections~\ref{substochastic_broadening} and \ref{subphase_diagram}.

\section{Results and discussion}
\label{results}

\subsection{Deterministic threshold and burnout boundary}
\label{subdeterministic_threshold}

The deterministic reference response is obtained by setting $\sigma_n=\sigma_\Theta=0$. 
In this limit, the stochastic equations reduce to the dimensionless carrier--thermal model given by Eqs.~\eqref{n_dimensionless} and \eqref{theta_dimensionless}.
Deterministic dynamics therefore define the reference boundary for stochastic threshold broadening.
The deterministic response is classified by the thermal indicator $\Lambda_{\rm th}$ defined in Eq.~\eqref{lambda_th_dimensionless}.
Trajectories with $\Lambda_{\rm th}<1$ are recoverable, while those with $\Lambda_{\rm th}\ge 1$ reach the deterministic burnout boundary.
The deposited work indicator $\Lambda_{\rm dep}$ measures accumulated electrothermal input, but heat removal prevents it from serving as a criterion for boundary crossing.

\begin{figure}[t]
\centering

\begin{subfigure}[t]{0.32\textwidth}
\centering
\includegraphics[width=\textwidth]{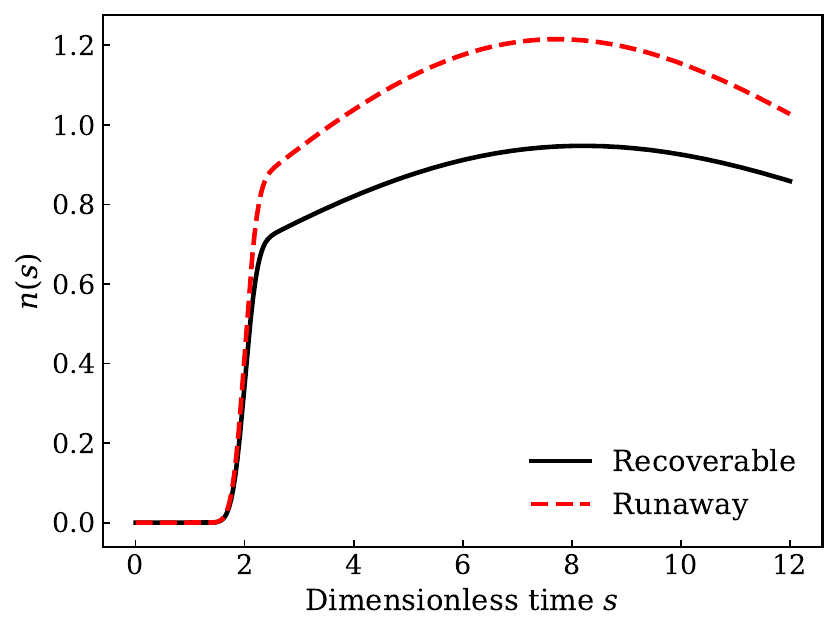}
\caption{}
\label{fig_deterministic_carrier}
\end{subfigure}
\hfill
\begin{subfigure}[t]{0.32\textwidth}
\centering
\includegraphics[width=\textwidth]{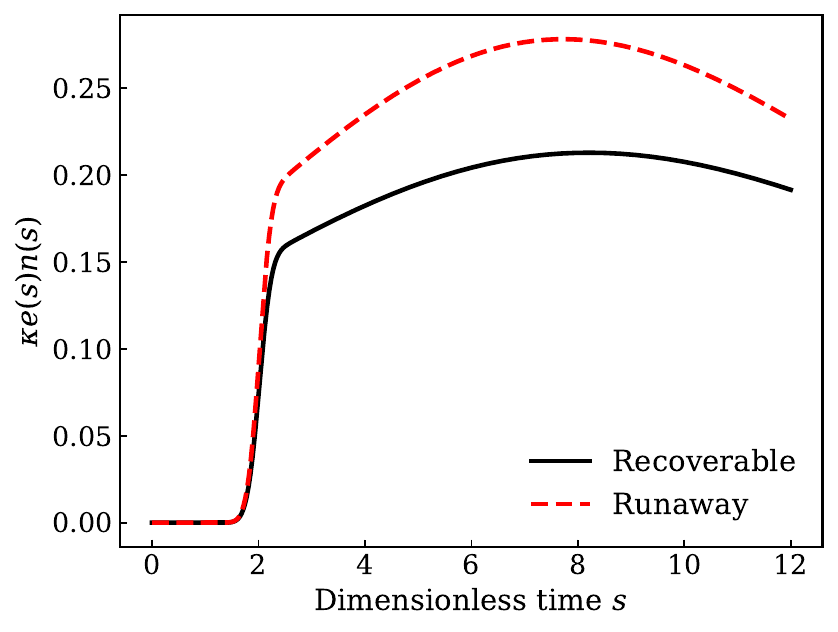}
\caption{}
\label{fig_deterministic_source}
\end{subfigure}
\hfill
\begin{subfigure}[t]{0.32\textwidth}
\centering
\includegraphics[width=\textwidth]{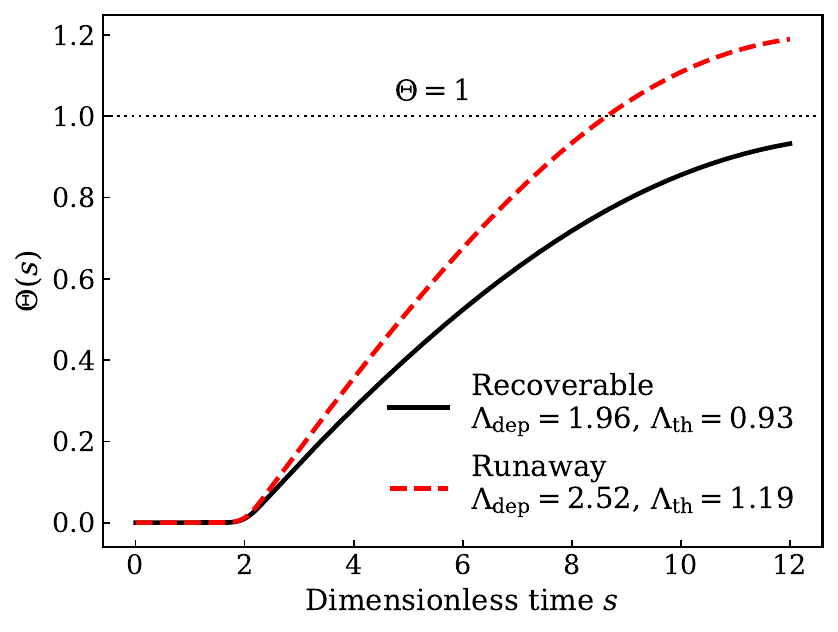}
\caption{}
\label{fig_deterministic_temperature}
\end{subfigure}
\caption{Deterministic reference trajectories of the reduced electrothermal model. 
(a) Carrier population $n(s)$. 
(b) Instantaneous electrothermal source $\kappa e(s)n(s)$. 
(c) Normalized temperature $\Theta(s)$.}
\label{fig_deterministic_traces}
\end{figure}

For the representative comparison in Figure~\ref{fig_deterministic_traces}, both cases are evaluated at the
same field value $b=1.20$, and differ only in ionization strength. 
The recoverable trajectory corresponds to $\ell=0.70$, whereas the runaway trajectory is $\ell=0.85$.
These values are chosen on opposite sides of the deterministic boundary.
In Figure~\ref{fig_deterministic_carrier}, the ionization pulse produces a transient carrier population for each trajectory.
The runaway trajectory reaches a larger carrier peak and decays more slowly after the pulse. 
This indicates that avalanche feedback sustains the carrier population more effectively for the larger ionization strength. 
The recoverable trajectory relaxes faster because carrier loss dominates the feedback process after the pulse.

Figure~\ref{fig_deterministic_source} shows the corresponding electrothermal source $\kappa e(s)n(s)$. 
The runaway case has a larger source amplitude and a more sustained heating interval. 
The recoverable case also receives finite electrothermal input, but the source decays before the temperature reaches the absorbing boundary. 
This difference shows that the temporal distribution of heating is important in addition to the total deposited work.

The corresponding normalized temperature response is shown in Fig.~\ref{fig_deterministic_temperature}. 
The recoverable trajectory remains below the absorbing boundary, with $\Lambda_{\rm dep}=1.96$ and $\Lambda_{\rm th}=0.93$. 
Although the accumulated electrothermal input exceeds the nominal thermal margin, thermal relaxation keeps the peak
temperature below the failure boundary. 
The runaway trajectory crosses the boundary, with $\Lambda_{\rm dep}=2.52$ and $\Lambda_{\rm th}=1.19$, because the thermal first-passage indicator exceeds unity. 
This comparison shows that $\Lambda_{\rm dep}>1$ is not a sufficient burnout criterion. 
Boundary crossing requires the thermal response to reach the absorbing boundary, rather than accumulated input alone.

The deterministic boundary in the $(\ell,b)$ plane is obtained from
\begin{equation}
\Lambda_{\rm th}(\ell,b)=1,
\label{deterministic_boundary_ell_b}
\end{equation}
with the remaining dimensionless parameters held fixed. 
Increasing $b$ raises the effective field and strengthens avalanche feedback, thereby reducing the critical ionization strength needed to reach $\Theta=1$.
Carrier loss and thermal relaxation shift the boundary in the opposite direction by favoring recovery. 
This boundary provides the deterministic reference curve for the subsequent stochastic transition analysis.

\subsection{Stochastic broadening of the burnout transition}
\label{substochastic_broadening}

In the zero-noise limit, the transition is controlled by the deterministic boundary $\Lambda_{\rm th}=1$. 
Finite fluctuations broaden this boundary into a smooth probability transition. 
The ionization strength $\ell$ is scanned across the deterministic transition region, with parameter $\sigma$ setting the noise amplitude through Eq.~\eqref{numerical_noise_convention}. 
The resulting $P_{\rm SEB}(\ell)$ curves give the transition width $\Delta\ell$. 
Near the deterministic threshold, trajectories with the same nominal parameters may either relax or reach the absorbing boundary. 
This variability reflects stochastic perturbations around the feedback-relaxation balance.

\begin{figure}[t]
\centering
\begin{subfigure}[t]{0.32\textwidth}
\centering
\includegraphics[width=\textwidth]{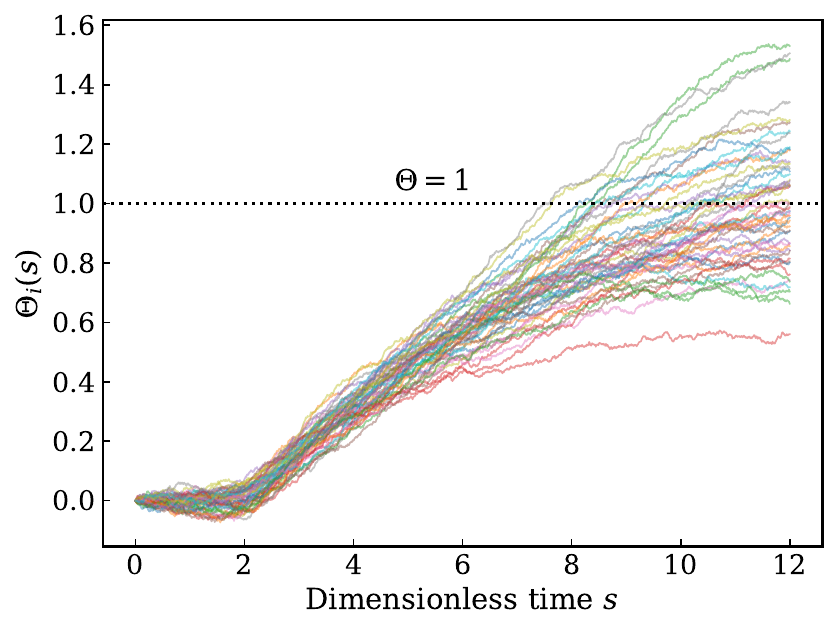}
\caption{}
\label{fig_stochastic_temperature_trajectories}
\end{subfigure}
\begin{subfigure}[t]{0.32\textwidth}
\centering
\includegraphics[width=\textwidth]{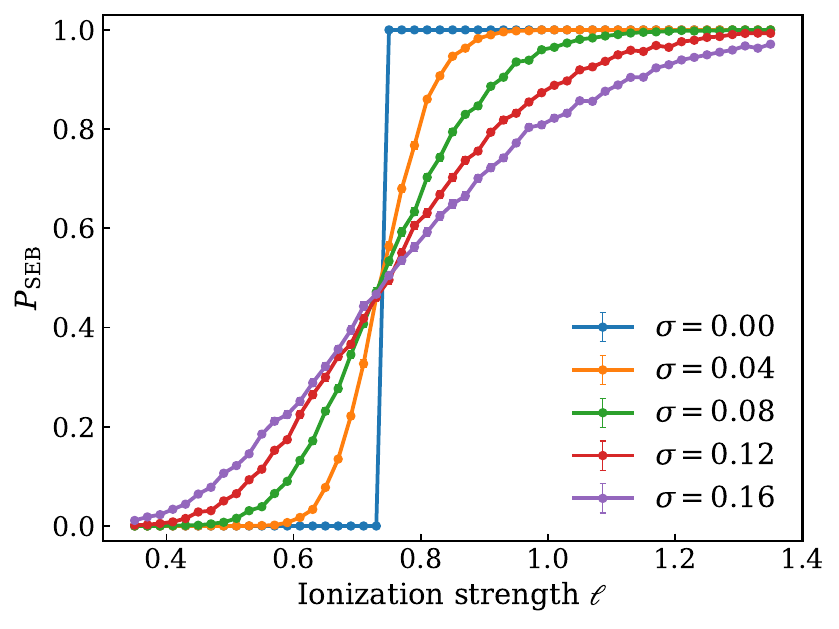}
\caption{}
\label{fig_stochastic_probability_curves}
\end{subfigure}
\begin{subfigure}[t]{0.32\textwidth}
\centering
\includegraphics[width=\textwidth]{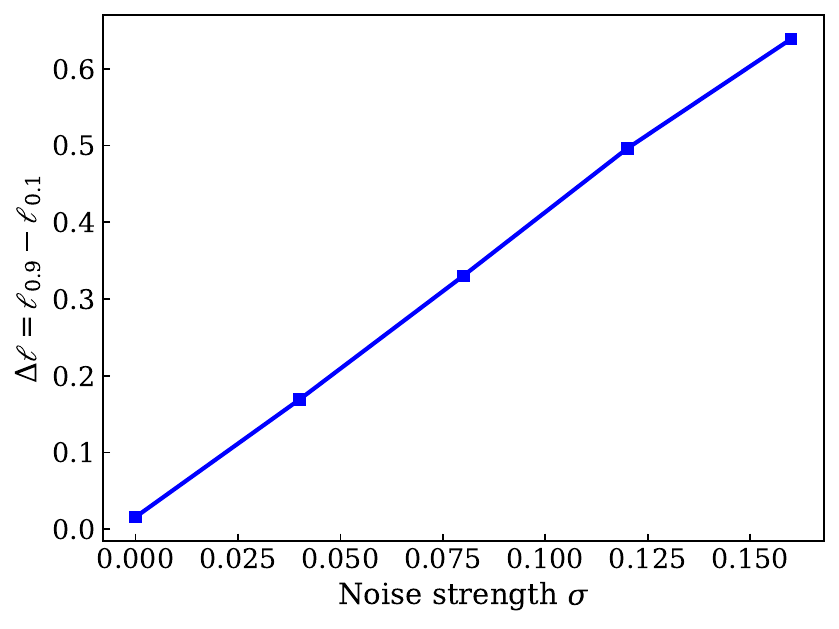}
\caption{}
\label{fig_stochastic_transition_width}
\end{subfigure}
\caption{Stochastic broadening of the burnout transition. (a)
Stochastic temperature trajectories near the deterministic boundary.
(b) Burnout probability as a function of ionization strength. Error
bars denote the binomial standard error. (c) Transition width as a
function of noise strength.}
\label{fig_stochastic_broadening}
\end{figure}

Figure~\ref{fig_stochastic_broadening} shows the stochastic broadening of the deterministic threshold. 
To illustrate the trajectory-level origin of this broadening, Figure~\ref{fig_stochastic_temperature_trajectories} shows temperature histories generated near the deterministic boundary.
Burnout is identified by crossing the absorbing boundary $\Theta=1$.
Although the trajectories share the same nominal control parameters, stochastic fluctuations produce different transient temperature
responses. 
Some realizations remain below the boundary, whereas others cross it within the observation window.
This coexistence occurs because feedback and relaxation are nearly balanced near the deterministic threshold, so small stochastic perturbations can change whether a trajectory reaches the absorbing boundary.

The probability curves in Figure~\ref{fig_stochastic_probability_curves} show the corresponding change in $P_{\rm SEB}$ as $\ell$ is increased.
For $\sigma=0$, the transition is sharp and reflects the deterministic threshold. 
As $\sigma$ increases, the transition becomes smoother and extends over a wider interval of $\ell$. 
Finite fluctuations allow burnout below the deterministic threshold and also delay the approach to near-certain burnout above the threshold. The error bars are small relative to the probability variation, which is consistent with the ensemble size used in the simulations. 
This behavior reflects stochastic spreading around the deterministic threshold, which permits subthreshold boundary crossing and delayed
saturation above the threshold.

The transition width in Figure~\ref{fig_stochastic_transition_width} increases with the noise strength.
This confirms that stronger fluctuations enlarge the interval in ionization strength over, which recoverable and runaway outcomes coexist. 
The trend should be interpreted as stochastic threshold broadening within the reduced model, rather than as a calibrated device-specific scaling law. This broadening arises because larger noise amplitudes produce stronger trajectory spreading around the feedback--relaxation balance.

The results show that a single deterministic threshold is insufficient to describe burnout near the transition. 
The condition $\Lambda_{\rm th}=1$ remains a useful reference boundary, but finite fluctuations convert it into a probability transition described by $P_{\rm SEB}(\ell)$. 
The finite transition width therefore quantifies the departure from a sharp deterministic threshold.

\subsection{Noise-induced subthreshold runaway}
\label{subsubthreshold}

The stochastic formulation also predicts runaway below the deterministic boundary. 
For the case considered here, $\Lambda_{\rm th}=0.93<1$, indicating a recoverable deterministic response. 
The deterministic trajectory therefore remains below the absorbing boundary during the observation window.
Finite fluctuations can nevertheless drive a subset of trajectories to $\Theta=1$. 
Such crossings arise from transient enhancement of carrier multiplication and local electrothermal heating.
The subthreshold stochastic regime is therefore characterized by
\begin{equation}
\Lambda_{\rm th}<1,
\qquad
P_{\rm SEB}>0 .
\label{subthreshold_probability_condition}
\end{equation}
This condition compares the deterministic reference with the stochastic ensemble response.
The numerical example fixes the ionization strength at $\ell=0.70$, which corresponds to the recoverable deterministic trajectory in
Fig.~\ref{fig_deterministic_traces}. 
Parameter $\sigma$ is varied according to Eq.~\eqref{numerical_noise_convention} to evaluate the probability of noise-activated burnout in this deterministically recoverable case.

\begin{figure}[t]
\centering
\begin{subfigure}[t]{0.32\textwidth}
\centering
\includegraphics[width=\textwidth]{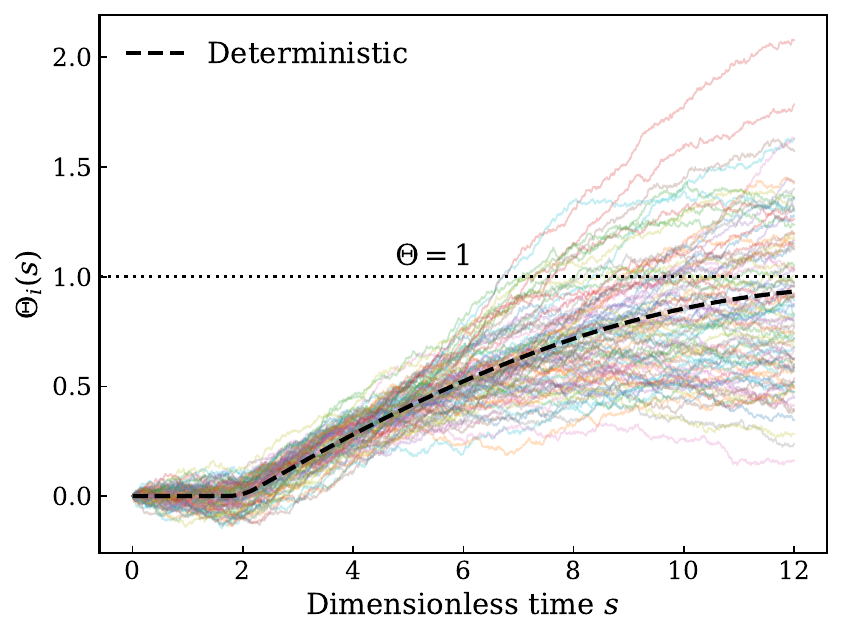}
\caption{}
\label{fig_subthreshold_stochastic_trajectories}
\end{subfigure}
\begin{subfigure}[t]{0.32\textwidth}
\centering
\includegraphics[width=\textwidth]{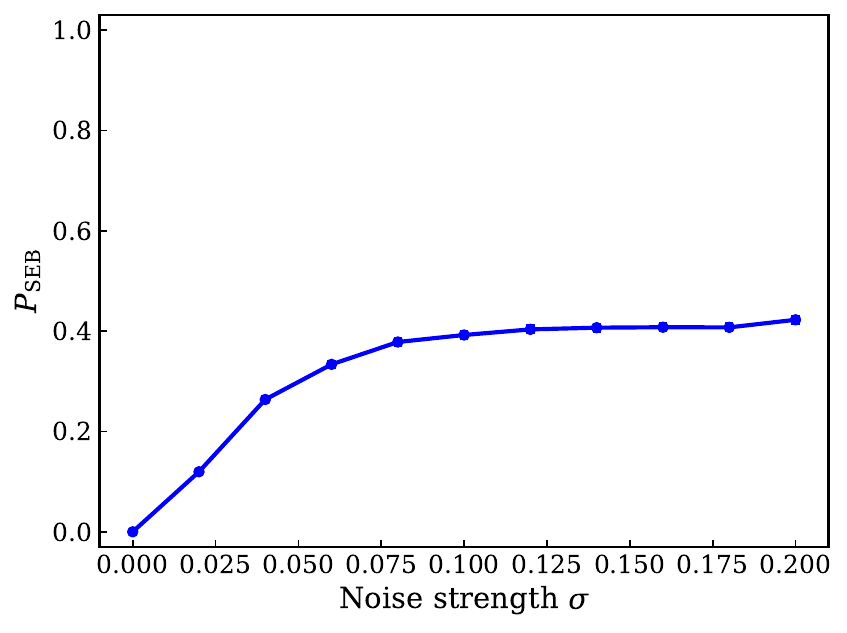}
\caption{}
\label{fig_subthreshold_probability_vs_noise}
\end{subfigure}
\begin{subfigure}[t]{0.32\textwidth}
\centering
\includegraphics[width=\textwidth]{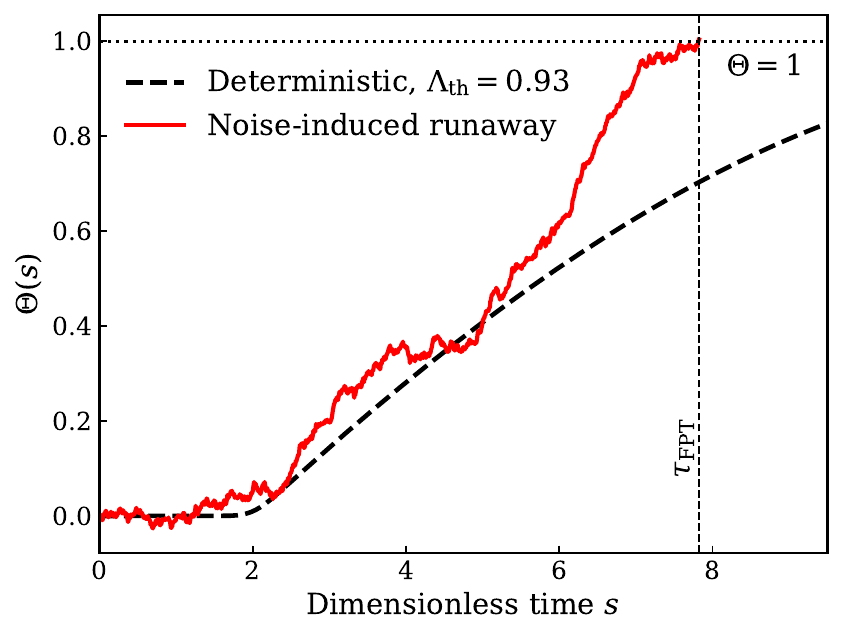}
\caption{}
\label{fig_subthreshold_rare_event_comparison}
\end{subfigure}
\caption{Noise-induced subthreshold electrothermal runaway. (a)
Stochastic temperature trajectories below the deterministic boundary.
(b) Burnout probability as a function of noise strength. Error bars
denote the binomial standard error. (c) Comparison between a
deterministic recoverable trajectory and a stochastic runaway event.}
\label{fig_subthreshold_runaway}
\end{figure}

Figure~\ref{fig_subthreshold_runaway} shows the subthreshold stochastic response. 
In Figure~\ref{fig_subthreshold_stochastic_trajectories}, the deterministic trajectory remains below $\Theta=1$, while the stochastic ensemble spreads around it. 
Most realizations follow the recoverable tendency, but a fraction of the trajectories crosses the absorbing boundary. This coexistence confirms that a deterministic subthreshold condition does not imply zero burnout probability once fluctuations are included.

Figure~\ref{fig_subthreshold_probability_vs_noise} shows the burnout probability as a function of the noise strength. At $\sigma=0$, the probability is zero within the MC estimate, consistent with the deterministic classification. 
As $\sigma$ increases, boundary-crossing events become more frequent and $P_{\rm SEB}$ rises. 
The curve does not represent a deterministic instability threshold. 
It measures the increasing probability of noise-activated first passage for a parameter set that remains recoverable in the noise-free model.
This behavior occurs because stronger fluctuations increase the chance that a recoverable trajectory makes a rare excursion to the absorbing thermal boundary.

Figure~\ref{fig_subthreshold_rare_event_comparison} compares the deterministic recoverable response with one representative noise-induced runaway trajectory. 
The deterministic trajectory rises after the ionization pulse and stays below the absorbing boundary. 
The stochastic trajectory follows a different temporal path and reaches $\Theta=1$ at the first-passage time
$\tau_{\rm FPT}$. 
The stochastic trajectory is plotted only up to this crossing, since the absorbing boundary represents irreversible failure in the reduced model.
This rare crossing results from a transient stochastic amplification of the electrothermal feedback before relaxation can restore the trajectory.

These results show that the deterministic condition $\Lambda_{\rm th}<1$ is a reference for the noise-free system, not an
absolute failure-free criterion. 
Below the deterministic boundary, stochastic amplification of electrothermal feedback can still drive rare crossings of the absorbing thermal boundary. 
The burnout margin is therefore statistical in the reduced stochastic description.

\subsection{First-passage-time statistics}
\label{subfpt}

\begin{figure}[t]
\centering
\begin{subfigure}[t]{0.32\textwidth}
\centering
\includegraphics[width=\textwidth]{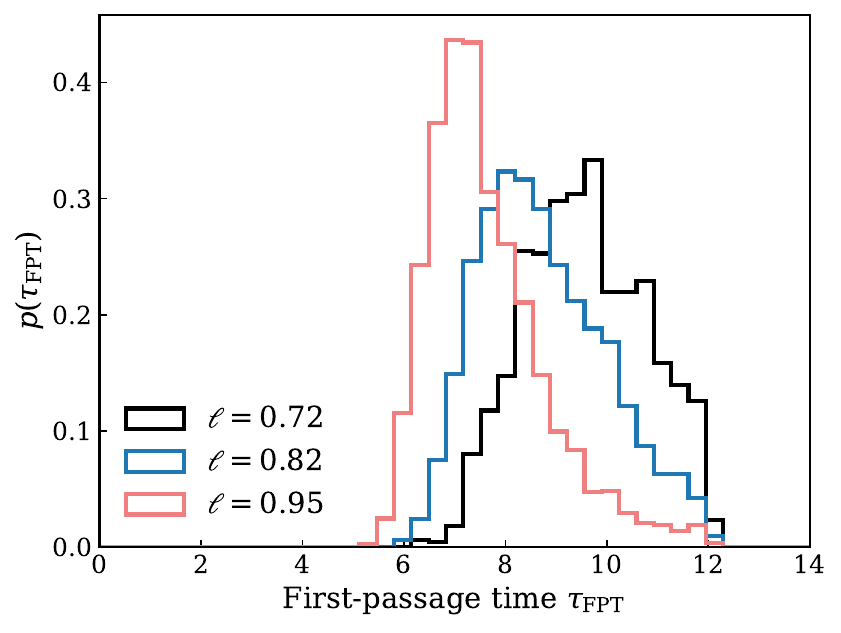}
\caption{}
\label{fig_fpt_distribution}
\end{subfigure}
\begin{subfigure}[t]{0.32\textwidth}
\centering
\includegraphics[width=\textwidth]{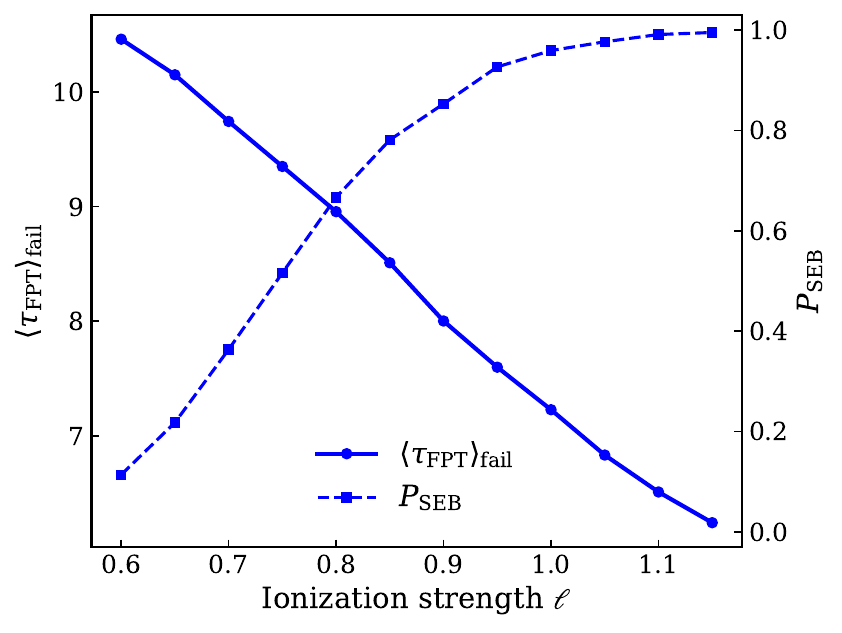}
\caption{}
\label{fig_fpt_mean_probability}
\end{subfigure}
\begin{subfigure}[t]{0.32\textwidth}
\centering
\includegraphics[width=\textwidth]{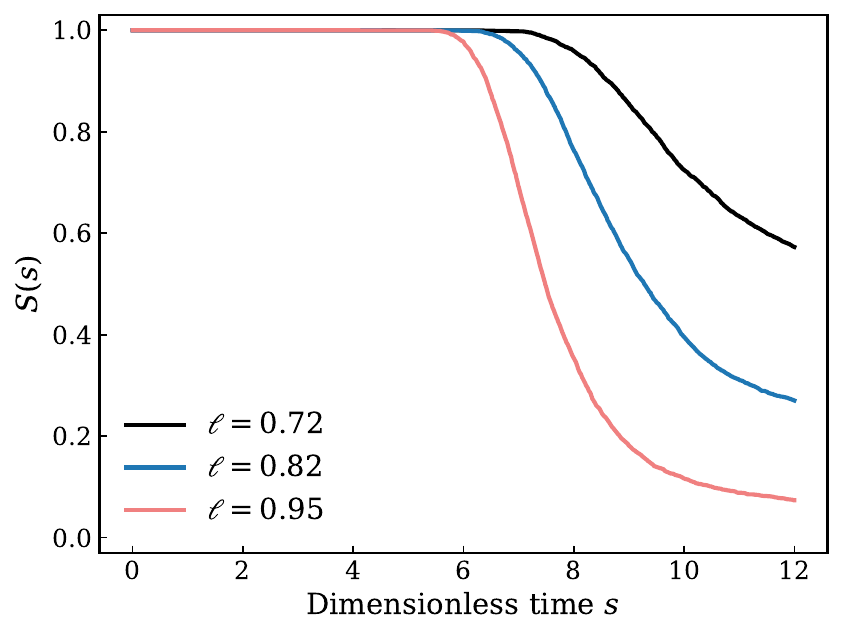}
\caption{}
\label{fig_fpt_survival_probability}
\end{subfigure}
\caption{First-passage-time statistics of stochastic electrothermal
runaway. (a) Distribution of first-passage times. (b) Conditional mean
first-passage time for failed trajectories and burnout probability as
functions of ionization strength. (c) Survival probability.}
\label{fig_fpt_statistics}
\end{figure}

The first-passage-time statistics characterize the temporal structure of stochastic electrothermal runaway. 
$P_{\rm SEB}$ gives the probability of boundary crossing within the observation window. 
The first-passage time $\tau_{\rm FPT}$ resolves the time scale of this crossing.
Figure~\ref{fig_fpt_statistics} shows first-passage-time statistics for representative ionization strengths selected from different parts of
the probabilistic transition band, with $\ell=0.72$, $\ell=0.82$, and $\ell=0.95$. 
This choice allows delayed stochastic failure and rapid runaway to be compared within the same reduced model. 
Near the lower part of the transition band, recovery and failure coexist under the same nominal parameters. 
At larger $\ell$, stronger carrier multiplication and electrothermal heating drive earlier boundary crossing. 
The conditional mean first-passage time, combined with $P_{\rm SEB}$, distinguishes rare delayed failure from rapid runaway.

In Fig.~\ref{fig_fpt_distribution}, the distribution for the smallest ionization strength is shifted toward later times and
is relatively broad. 
This behavior corresponds to delayed noise-assisted failure, where the deterministic drift remains close to the recoverable regime and crossing requires a favorable stochastic excursion.
As $\ell$ increases, the distributions move toward earlier times and become more concentrated, indicating faster boundary crossing
under stronger electrothermal feedback.

The scan in Fig.~\ref{fig_fpt_mean_probability} separates the timing and frequency of failure. 
The conditional mean first-passage time decreases as $\ell$ increases.
Failed trajectories therefore reach the absorbing boundary earlier under stronger ionization drive.
Over the same range, $P_{\rm SEB}$ increases. 
Thus, increasing $\ell$ raises the likelihood of burnout and shortens the conditional failure time among failed trajectories.
This trend arises because stronger ionization increases carrier growth and electrothermal feedback, reducing the time needed to reach the
absorbing boundary.

The survival curves in Fig.~\ref{fig_fpt_survival_probability} provide a time-resolved representation of the same first-passage process. 
For smaller $\ell$, $S(s)$ decays slowly, indicating that many trajectories remain below the absorbing boundary during the observation window. 
For larger $\ell$, the survival probability decreases more rapidly, showing that boundary crossing occurs earlier and in a larger fraction
of the ensemble.
This faster decay reflects the stronger drift toward the absorbing thermal boundary produced by enhanced carrier multiplication and
electrothermal heating.

These first-passage observables complement the burnout probability. 
$P_{\rm SEB}$ measures the occurrence of burnout within the observation window. 
In contrast, $p(\tau_{\rm FPT})$, $\langle\tau_{\rm FPT}\rangle_{\rm fail}$, and $S(s)$ resolve the temporal structure of boundary crossing. This distinction is important near the stochastic transition band, where rare delayed failure and rapid runaway can coexist within the same coarse-grained electrothermal dynamics.

\subsection{Feedback--relaxation phase diagram}
\label{subphase_diagram}

\begin{figure}[t]
\centering
\includegraphics[width=0.6\textwidth]{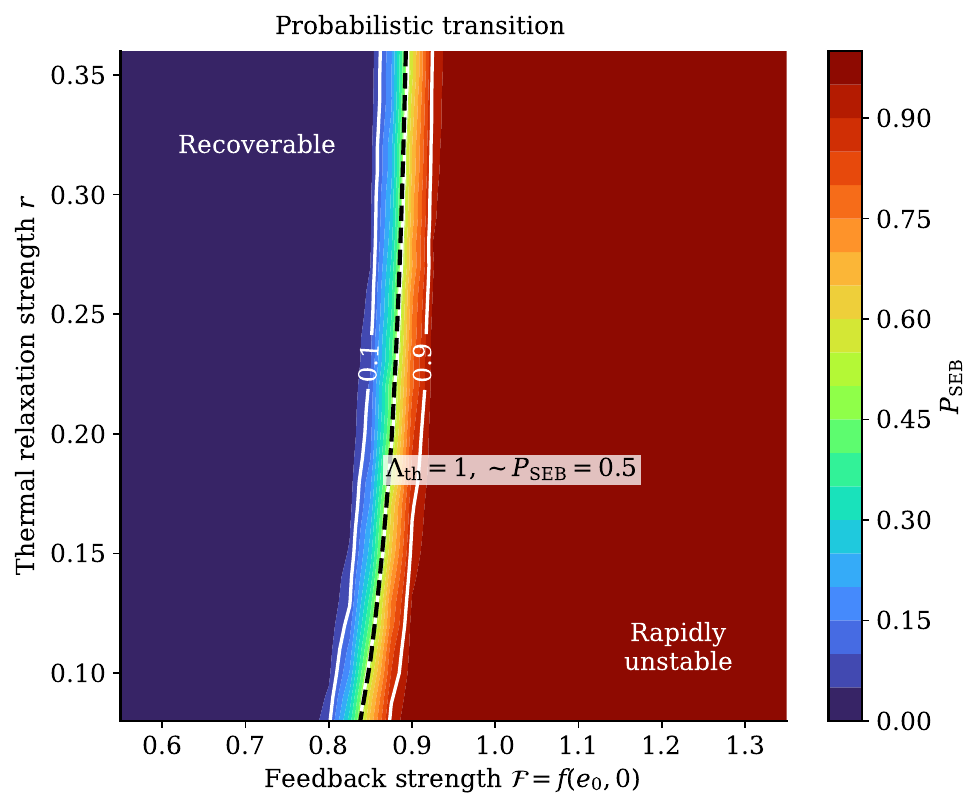}
\caption{Feedback--relaxation phase diagram. The color map shows $P_{\rm SEB}$ in the $(\mathcal F,r)$ plane. The contours
$P_{\rm SEB}=0.1$, $0.5$, and $0.9$ mark the stochastic transition band. The dashed contour denotes the deterministic
reference boundary $\Lambda_{\rm th}=1$.}
\label{fig_phase_diagram}
\end{figure}

Figure~\ref{fig_phase_diagram} summarizes the stochastic threshold
behavior of the reduced model in the feedback--relaxation plane. The
feedback coordinate is defined by the reference avalanche factor
\begin{equation}
\mathcal F=f(e_0,0),
\label{feedback_strength_dimensionless}
\end{equation}
where $e_0$ is the reference dimensionless field. Larger $\mathcal F$
corresponds to stronger avalanche multiplication, while larger $r$
represents stronger thermal relaxation.
The phase diagram is obtained by evaluating $P_{\rm SEB}$ on the $(\mathcal F,r)$ plane with the remaining dimensionless parameters held
fixed. 
The coordinate $\mathcal F$ is varied through the avalanche scale $A_f$, while $B_f$ and $a_\Theta$ are kept fixed.
This construction isolates the competition between avalanche feedback and thermal recovery in the reduced stochastic model.

The probability map separates recoverable, probabilistic, and rapidly unstable regimes. 
Regions with small $P_{\rm SEB}$ correspond to trajectories that usually remain below the absorbing boundary. 
Regions with large $P_{\rm SEB}$ correspond to boundary crossing within the observation window for most trajectories. 
Increasing $\mathcal F$ strengthens avalanche multiplication and electrothermal heating, which shifts the system toward burnout. 
Stronger thermal relaxation, represented by a larger $r$, shifts the transition toward larger feedback strength at fixed burnout probability.

The probability contours show that finite fluctuations replace the deterministic boundary by a transition band. 
Within this band, trajectories with the same nominal control parameters can either relax or reach $\Theta=1$. 
The dashed contour $\Lambda_{\rm th}=1$ lies close to the middle of the stochastic transition, near $P_{\rm SEB}=0.5$.
The finite spacing between the $P_{\rm SEB}=0.1$ and $P_{\rm SEB}=0.9$ contours shows the broadening of the deterministic
threshold in the $(\mathcal F,r)$ plane.

This representation gives a compact statistical interpretation of device-scale changes. 
In SiC power MOSFETs, stronger field enhancement or stronger overlap between the ion track and the high-field region can
be represented by an increase in $\mathcal F$. 
More efficient thermal spreading can be represented by an increase in $r$. 
Quantitative mapping of a particular device onto this diagram would require calibration of the effective parameters against TCAD simulations or irradiation data.

\section{Discussion}\label{discussion}

\subsection{Connection to SEB in SiC power MOSFETs}
\label{sic_mapping}

\begin{figure}[t]
\centering
\includegraphics[width=0.98\textwidth]{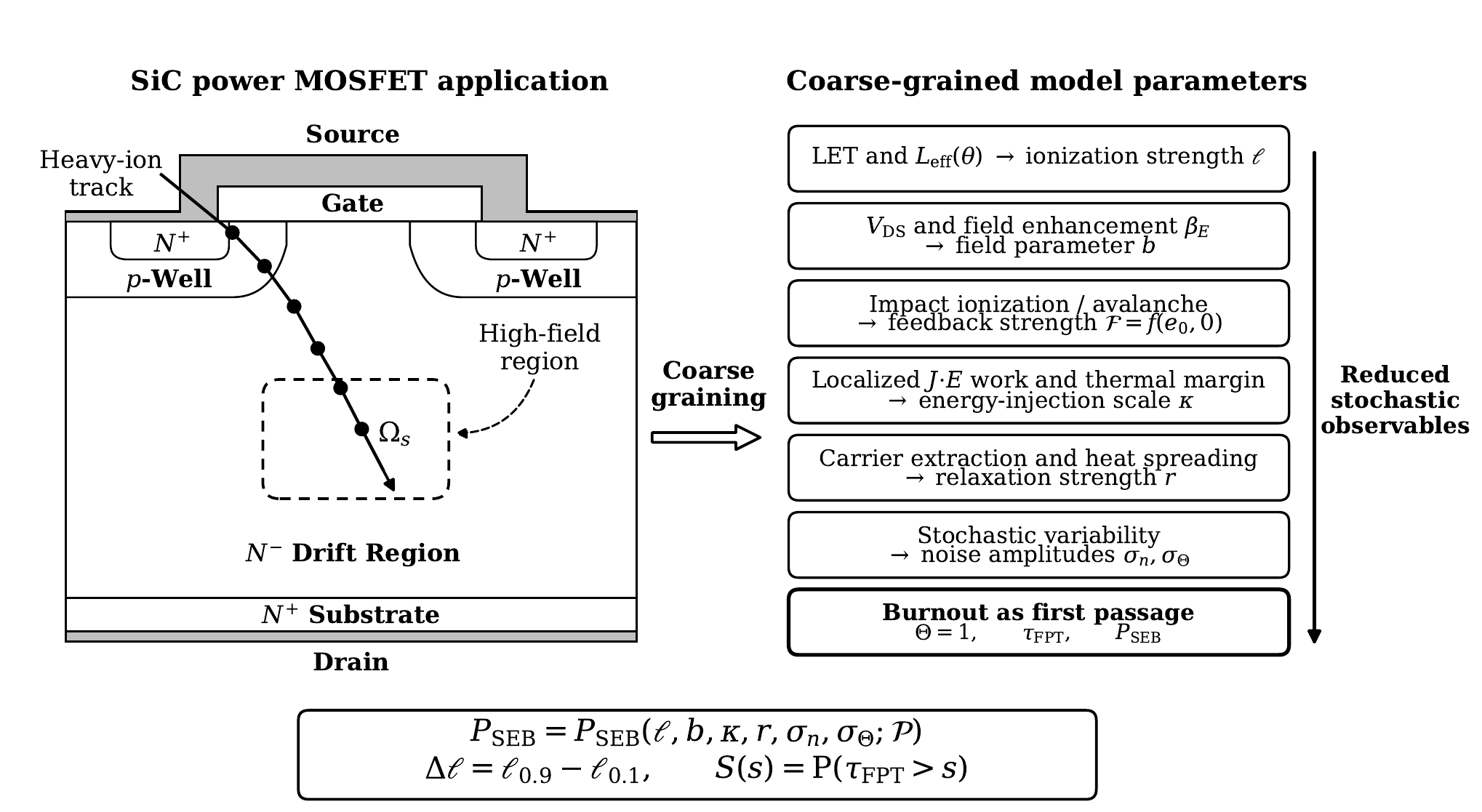}
\caption{
Coarse-grained mapping from SiC power-device SEB physics to the
reduced stochastic electrothermal model.
}
\label{fig_sic_mapping}
\end{figure}

The reduced stochastic model can be interpreted as a coarse-grained description of heavy-ion-induced SEB in SiC power MOSFETs. 
This mapping relates the statistical variables of the reduced model to the main physical ingredients of the device response.
Figure~\ref{fig_sic_mapping} summarizes this connection.
The left panel shows the device picture of a vertical SiC power MOSFET, which conducts between the top source contact
and the bottom drain contact. 
Under off-state drain bias, the lightly doped $N^{-}$ drift region supports most of the blocking voltage. 
High electric fields can develop in localized field-crowding regions near the gate and drift region \cite{Wang2024,Germanicus2025,Liao2024}.
A heavy-ion track passing through or near this high-field region generates
excess carriers. 
These carriers may be collected, multiplied by impact ionization, and coupled to localized Joule heating \cite{Grome2024,Mo2023,WangNegGate2024}.

The reduced model does not resolve the full spatial structure of the MOSFET. 
Instead, the right panel of Figure~\ref{fig_sic_mapping} maps the device response onto collective variables, effective parameters, and
first-passage observables. 
The sensitive region $\Omega_s$ represents the part of the depleted drift region where the local electric field,
ion-track charge deposition, and electrothermal heating are most strongly coupled \cite{McPherson2020,Wang2024,Zhang2022,Germanicus2025}. 
A heavy-ion strike creates carriers in or near this region. 
Subsequent avalanche multiplication and local $\mathbf J\cdot\mathbf E$ work drive the temperature rise represented by
$\Theta(s)$, while boundary crossing at $\Theta=1$ defines burnout in the reduced stochastic model.

The ionization strength $\ell$ quantifies the effective primary charge injected into $\Omega_s$. 
In a SiC power MOSFET, $\ell$ increases with LET, ion-track overlap with the high-field region, and favorable strike geometry. 
A larger $\ell$ raises the carrier source and moves the reduced system toward the stochastic transition band. 
The field parameter $b$ sets the imposed local field scale. 
It is influenced by drain bias, drift-region design, junction curvature, and field crowding near sensitive locations. 
Larger values of $b$ strengthen field-assisted multiplication and increase the likelihood of reaching the absorbing boundary.

The feedback coordinate $\mathcal F$ characterizes the effective avalanche gain under a reference electrothermal condition. 
Within the SiC power-device setting, larger $\mathcal F$ corresponds to stronger local electric-field enhancement and impact ionization in the high-field region. 
The energy injection scale $\kappa$ measures the conversion of localized electrical work into normalized temperature rise. 
It reflects current localization and the thermal margin in the active device region.
The relaxation strength $r$ represents thermal recovery and carrier removal. 
An increase in $r$ represents more effective heat spreading or carrier extraction and shifts the response toward recovery.

The stochastic amplitudes $\sigma_n$ and $\sigma_\Theta$ capture event-level variability not explicitly resolved by the reduced
variables. 
For SiC power MOSFETs, this variability reflects fluctuations in ion-track charge deposition, avalanche paths near the high-field
region, and local thermal response. 
The absorbing boundary $\Theta=1$ represents the coarse-grained burnout condition.
The observables $P_{\rm SEB}$, $\Delta\ell$, $\tau_{\rm FPT}$, and $S(s)$ quantify the probability, width, timing, and survival characteristics of the failure process.

This correspondence also gives the expected SEB trends. 
Stronger ionization input, local field enhancement, or avalanche feedback shifts the reduced system toward burnout. 
More effective carrier extraction and heat spreading shift the response toward recovery. 
Higher ambient temperature acts in the damaging direction by reducing the available thermal margin. 
The deterministic boundary $\Lambda_{\rm th}=1$ remains a useful reference, but finite fluctuations turn the response into a
probability transition rather than a single critical LET or bias value.
In this form, SEB in SiC power MOSFETs is interpreted as electrothermal escape driven by ionization and field-assisted feedback.
Device-specific prediction still requires calibration against TCAD simulations or irradiation data.

\subsection{Difference from deterministic TCAD threshold analysis}
\label{tcad_difference}

Deterministic TCAD analysis and the present reduced stochastic model address different levels of the SEB problem. 
TCAD resolves the device geometry and doping profile under prescribed bias, temperature, and ion-strike conditions. 
It does so through coupled transport, electrostatic, and thermal equations \cite{Peng2021, McPherson2020, Zhang2022}.
Such a description is well suited to identifying spatial localization of electric field, current flow, and lattice temperature, as well as
geometry-dependent failure precursors.
The reduced model does not replace this device-resolved description.
Instead, the model projects the electrothermal response of a sensitive region onto the coarse-grained variables $n(s)$ and $\Theta(s)$.
Effective parameters such as $\ell$, $b$, $\mathcal F$, $\kappa$, and $r$ summarize ionization input, avalanche feedback, localized heating, and relaxation. 
The model is therefore a statistical first-passage description of threshold activation rather than a spatial simulator of a particular SiC MOSFET.

The main distinction lies in the interpretation of the burnout threshold. 
In deterministic TCAD, a fixed geometry and strike condition usually lead to a recoverable or failed trajectory at a prescribed bias
and LET \cite{Peng2021,McPherson2020,Zhang2022,Liao2024}. 
In the reduced stochastic formulation, the deterministic condition $\Lambda_{\rm th}=1$ serves as a reference boundary. 
Once finite fluctuations are included, the transition is described by $P_{\rm SEB}$ rather than by a single critical point.

This difference becomes important near the boundary between recovery and runaway. 
Deterministic TCAD can locate the nominal condition where electrothermal runaway appears for a given strike scenario. 
The stochastic model describes the dispersion around this nominal condition.
Fluctuations in ionization deposition, avalanche multiplication, local field, and thermal response can change the trajectory outcome. 
As a result, the same coarse-grained control parameters may lead to recovery or burnout.
The deterministic boundary is then broadened into a probability band, and $\Delta\ell$ measures the corresponding threshold dispersion.

The treatment of subthreshold response is also different. 
A deterministic calculation classifies a parameter set with $\Lambda_{\rm th}<1$ as recoverable for the specified trajectory. 
In the stochastic formulation, the same reference condition can coexist with a finite burnout probability when rare fluctuations drive the
trajectory to the absorbing boundary $\Theta=1$. 
This noise-induced subthreshold runaway is not inconsistent with deterministic TCAD. 
It indicates that the deterministic trajectory represents the nominal response, while stochastic variability determines the probability of
rare first-passage events around that response.

The two approaches are therefore complementary. 
TCAD provides spatially resolved electrothermal information, including distributions of electric field, current density, and lattice temperature. The reduced stochastic model converts feedback and relaxation mechanisms into probability and first-passage observables, such as $P_{\rm SEB}$, $\Delta\ell$, $p(\tau_{\rm FPT})$, $\langle\tau_{\rm FPT}\rangle_{\rm fail}$, and $S(s)$. 
Detailed limitations, calibration requirements, and possible extensions are discussed in Section~\ref{subdiscussion_limitations}.

\subsection{Numerical convergence}
\label{subnumerical_convergence}

\begin{figure}[t]
\centering
\includegraphics[width=0.72\textwidth]{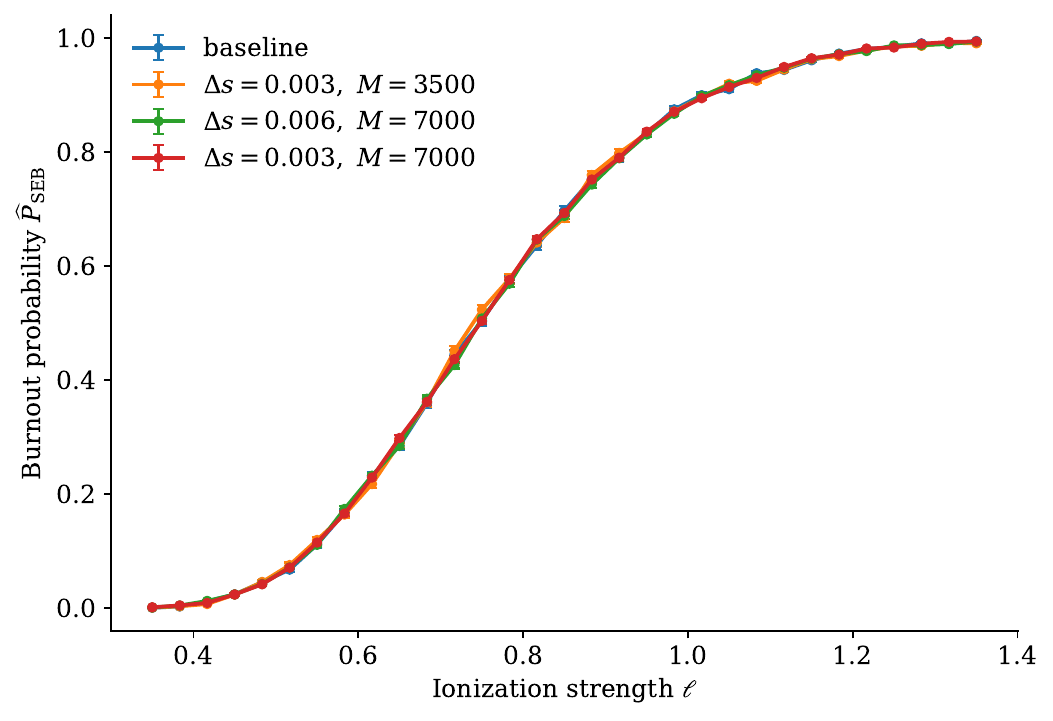}
\caption{Numerical convergence of the burnout probability curve under
time-step and ensemble-size refinement.}
\label{fig_numerical_convergence}
\end{figure}

The stochastic calculations use a finite time step and a finite trajectory ensemble. 
Starting from the baseline setting $\Delta s=0.006$ and $M=3500$, convergence is assessed by three refinements. 
The refinements are a reduced time step $\Delta s=0.003$, an increased ensemble size $M=7000$, and the combined refinement.
Figure~\ref{fig_numerical_convergence} shows the resulting $\widehat P_{\rm SEB}(\ell)$ curves. 
The baseline curve and the three refined curves nearly overlap over the scanned range of ionization strength.
Neither time-step refinement nor ensemble-size refinement produces a systematic shift of the transition region. 
This agreement is most relevant in the intermediate probability range, where the transition width is extracted.

The convergence of $\Delta\ell$ is evaluated from the same probability curves using the interpolation procedure described above.
The baseline calculation gives $\Delta\ell=0.4773$, while the refined calculations yield values from $0.4794$ to $0.4881$.
Across the refined calculations, the variation is about $1.1\times10^{-2}$, corresponding to approximately $2.3\%$ of the
baseline value.
At the present numerical resolution, the extracted stochastic broadening is therefore insensitive to the time step and ensemble size.

Residual differences among the probability curves are comparable to the MC sampling scale.
From Eq.~\eqref{p_standard_error}, the maximum binomial standard error for $M=3500$ occurs near $\widehat P_{\rm SEB}=0.5$ and is approximately $8.5\times10^{-3}$.
The residual curve-to-curve variations in Figure~\ref{fig_numerical_convergence} are of this order. 
Thus, the uncertainty in the transition band is dominated by finite ensemble sampling rather than by Euler-Maruyama time discretization.

Numerical boundary projections are also monitored.
For the tested parameter values, neither the carrier projection enforcing $n\ge0$ nor the lower thermal projection at $\Theta_{\min}$ becomes active. 
These projections therefore do not affect the reported burnout probabilities, transition widths, or first-passage-time statistics. 
The absorbing boundary at $\Theta=1$ remains the active boundary that determines the first-passage events.

The convergence results support the use of $\Delta s=0.006$ and $M=3500$ for the stochastic calculations. 
Under refinement of the time step and ensemble size, the probability curves and transition widths remain stable.
Residual uncertainty is mainly associated with MC sampling near the middle of the probabilistic transition band, rather than with
time-step resolution or numerical boundary projections.

\subsection{Model limitations and possible extensions}
\label{subdiscussion_limitations}

The present model is intentionally reduced. Its purpose is to isolate the stochastic feedback-relaxation mechanism of electrothermal runaway, rather than to reproduce the full spatial response of a specific SiC power device. 
Under this approximation, the device geometry is replaced by an effective volume $\Omega_s$. 
Its electrothermal state is then represented by the collective variables $n(s)$ and $\Theta(s)$.
As a result, spatial distributions of electric field, current density, carrier density, and lattice temperature are not resolved. 
The model also does not predict the precise hot-spot location or the morphology of post-burnout damage.

Several physical mechanisms are outside this coarse-grained description. 
Detailed ion-track structure and microscopic charge-deposition statistics are not resolved explicitly. 
Other failure channels, such as parasitic bipolar triggering, oxide rupture, gate degradation, and metallization failure, are not treated as separate processes.
Their leading effects can only enter through effective parameters such as $\ell$, $\mathcal F$, $b$, $\kappa$, $r$, $\sigma_n$, and $\sigma_\Theta$. 
The absorbing boundary $\Theta=1$ should therefore be interpreted as an effective failure criterion, not as a detailed material-damage model.

The thermal and stochastic descriptions are also simplified. 
The variable $\Theta(s)$ represents an effective active-region temperature, so spatial heat diffusion and heat spreading into neighboring layers are not resolved. 
The stochastic terms are phenomenological. Carrier noise accounts for unresolved variability in avalanche growth, whereas thermal noise
accounts for unresolved thermal fluctuations.
The baseline calculations use independent noise sources, and the convention in Eq.~\eqref{numerical_noise_convention} should
not be interpreted as a measured device noise spectrum.

These limitations suggest several extensions. Correlated noise could represent coupled fluctuations in avalanche multiplication and local
heating, following standard stochastic-process extensions with  non-diagonal noise covariance \cite{Gardiner2009,VanKampen2007,Risken1996}.
A spatial heat-transport model could replace the lumped thermal equation when substrate thickness, buffer layers, metallization, or
thermal boundary resistance are important. 
The Gaussian ionization pulse could also be replaced by a source sampled from MC ion-transport calculations \cite{Ziegler2010,Agostinelli2003}. 
This extension would allow track geometry and sensitive-volume overlap to enter the reduced dynamics.

Quantitative use of the reduced model requires calibration and validation. 
TCAD calculations can provide local field enhancement, carrier removal times, and impact ionization rates. 
They can also estimate thermal resistance, thermal capacitance, and energy localization coefficients \cite{Peng2021,Germanicus2025, Liao2024}.
Heavy-ion SEB data can then be used to calibrate the probability transition curve and estimate the transition width \cite{Peng2021,Pocaterra2023}. 
The transition width $\Delta\ell$ is therefore interpreted as a finite-window measure of stochastic broadening, not as a universal scaling quantity.
Experimental data can also test the organization of different device designs on a common feedback-relaxation diagram after normalization.

In this sense, the present study should be viewed as an initial step toward a broader
statistical-physics framework for SEB in SiC power devices, especially SiC power MOSFETs. 
Future work will combine the reduced first-passage model with device-resolved TCAD simulations and heavy-ion irradiation
experiments.
Continued work along this direction would progressively refine the effective parameters of the reduced model and help bridge
statistical-physics theory with practical power-device reliability analysis.

\section{Conclusion}
\label{conclusion}

This work developed a reduced stochastic electrothermal model for heavy-ion-induced SEB in SiC power MOSFETs. 
The burnout process was formulated as a first-passage problem. 
In this formulation, the state space is spanned by the effective carrier population $n(s)$ and the normalized temperature $\Theta(s)$.
The dimensionless parameters summarize ionization input, electrothermal feedback, and relaxation. 
Burnout was represented by an absorbing thermal boundary at $\Theta=1$.

For the stochastic formulation, the deterministic part of the model serves as a reference response.
The thermal indicator $\Lambda_{\rm th}$ defines the deterministic boundary through $\Lambda_{\rm th}=1$, while the deposited work indicator $\Lambda_{\rm dep}$ measures accumulatedelectrothermal input. 
Their comparison shows that accumulated input alone is not a sufficient burnout criterion, because boundary crossing
depends on the thermal response after feedback and relaxation have acted.

Against this deterministic reference, finite fluctuations broaden the boundary into a probabilistic transition.
The burnout probability $P_{\rm SEB}(\ell)$ increases smoothly across the transition region. 
In this transition, $\Delta\ell$ quantifies the coexistence range of recoverable and runaway trajectories.
Noise-induced subthreshold burnout is also predicted by the model.
Even when $\Lambda_{\rm th}<1$, rare stochastic excursions can still drive trajectories to the absorbing boundary.

Beyond the burnout probability, first-passage observables characterize the temporal structure of boundary crossing.
Using $p(\tau_{\rm FPT})$,  $\langle\tau_{\rm FPT}\rangle_{\rm fail}$, and $S(s)$, the model distinguishes rapid runaway from delayed stochastic failure.
The feedback-relaxation phase diagram further organizes the response in the $(\mathcal F,r)$ plane. Recoverable, probabilistic, and rapidly
unstable regimes are separated by the probability contours.

The resulting framework provides a statistical interpretation of threshold dispersion in SEB of SiC power MOSFETs. 
As a complement to deterministic TCAD analysis and irradiation testing, the framework links coarse-grained electrothermal dynamics to probability and first-passage observables.
Quantitative prediction for a specific device still requires calibration against device-resolved simulations and irradiation data. 
Further work will combine the reduced first-passage formulation with device-resolved simulations and irradiation data to refine the effective
parameters and test the predicted threshold broadening.

\section*{Acknowledgement}
Feiyi Liu, Min Guo and Mingyang Liu were supported by National Natural Science Foundation of China (Grant No.12564032), Yunnan Provincial Xiao Rui Expert Workstation (Grant No.202605AF350031), Independent Research Fund of Yunnan Tin \& Indium Laboratory (Grant No.202405AR340002-25BC01), Yunnan Provincial Department of Education Science Research Fund Project (Grant No.2025J0942, 2026J0977),  2025 Self-funded Science and Technology Projects of Chuxiong Prefecture (Grant No.cxzc2025004, cxzc2025008) and Chuxiong Normal University Doctoral Research Initiation Fund Project (Grant No.BSQD2407, BSQD2507). Shiyang Chen was supported by the China Scholarship Council (No.~202308420042) and Swansea University joint PhD project. Yang Wang was supported by Dongying Science Development Fund(Grant No.DJB2023015).

%\appendix

\bibliographystyle{elsarticle-num}
\bibliography{bibtex}

@article{Hardy1982,
  author  = {Hardy, Robert J.},
  title   = {Formulas for Determining Local Properties in Molecular-Dynamics Simulations: Shock Waves},
  journal = {The Journal of Chemical Physics},
  volume  = {76},
  number  = {1},
  pages   = {622--628},
  year    = {1982},
  doi     = {10.1063/1.442714}
}

@book{Gardiner2009,
  author    = {Gardiner, Crispin W.},
  title     = {Stochastic Methods: A Handbook for the Natural and Social Sciences},
  edition   = {4},
  publisher = {Springer},
  address   = {Berlin},
  year      = {2009},
  isbn      = {978-3-540-70712-7}
}

@book{VanKampen2007,
  author    = {van Kampen, Nicolaas G.},
  title     = {Stochastic Processes in Physics and Chemistry},
  edition   = {3},
  publisher = {North-Holland},
  address   = {Amsterdam},
  year      = {2007},
  isbn      = {978-0-444-52965-7}
}

@book{Risken1996,
  author    = {Risken, Hannes},
  title     = {The Fokker--Planck Equation: Methods of Solution and Applications},
  edition   = {2},
  publisher = {Springer},
  address   = {Berlin},
  year      = {1996},
  doi       = {10.1007/978-3-642-61544-3},
  isbn      = {978-3-642-61544-3}
}

@book{Redner2001,
  author    = {Redner, Sidney},
  title     = {A Guide to First-Passage Processes},
  publisher = {Cambridge University Press},
  address   = {Cambridge},
  year      = {2001},
  doi       = {10.1017/CBO9780511606014},
  isbn      = {978-0-521-65248-3}
}

@book{Baliga2008,
  author    = {Baliga, B. Jayant},
  title     = {Fundamentals of Power Semiconductor Devices},
  publisher = {Springer},
  address   = {New York},
  year      = {2008},
  doi       = {10.1007/978-0-387-47314-7},
  isbn      = {978-0-387-47314-7}
}

@book{Kimoto2014,
  author    = {Kimoto, Tsunenobu and Cooper, James A.},
  title     = {Fundamentals of Silicon Carbide Technology: Growth, Characterization, Devices and Applications},
  publisher = {Wiley},
  address   = {Singapore},
  year      = {2014},
  doi       = {10.1002/9781118313534},
  isbn      = {978-1-118-31352-7}
}

@book{Messenger1992,
  author    = {Messenger, George C. and Ash, Milton S.},
  title     = {The Effects of Radiation on Electronic Systems},
  edition   = {2},
  publisher = {Van Nostrand Reinhold},
  address   = {New York},
  year      = {1992},
  isbn      = {978-0-442-23952-7}
}

@book{Petersen2011,
  author    = {Petersen, Edward L.},
  title     = {Single Event Effects in Aerospace},
  publisher = {Wiley},
  address   = {Hoboken},
  year      = {2011},
  doi       = {10.1002/9781118084328},
  isbn      = {978-0-470-76749-9}
}

@article{Titus2013,
  author  = {Titus, Jeffrey L.},
  title   = {An Updated Perspective of Single Event Gate Rupture and Single Event Burnout in Power MOSFETs},
  journal = {IEEE Transactions on Nuclear Science},
  volume  = {60},
  number  = {3},
  pages   = {1912--1928},
  year    = {2013},
  doi     = {10.1109/TNS.2013.2252194}
}

@article{Zhang2022,
  author  = {Zhang, Hong and Guo, Hong-Xia and Zhang, Feng-Qi and Pan, Xiao-Yu and Liu, Yi-Tian and Gu, Zhao-Qiao and Ju, An-An and Ouyang, Xiao-Ping},
  title   = {Sensitivity of Heavy-Ion-Induced Single Event Burnout in SiC MOSFET},
  journal = {Chinese Physics B},
  volume  = {31},
  number  = {1},
  pages   = {018501},
  year    = {2022},
  doi     = {10.1088/1674-1056/ac051d}
}

@article{Kim2022,
  author  = {Kim, Junghun and Kim, Kwangsoo},
  title   = {Single-Event Burnout Hardening 4H-SiC UMOSFET Structure},
  journal = {IEEE Transactions on Device and Materials Reliability},
  volume  = {22},
  number  = {2},
  pages   = {164--168},
  year    = {2022},
  doi     = {10.1109/TDMR.2022.3151704}
}

@book{Sze2006,
  author    = {Sze, S. M. and Ng, Kwok K.},
  title     = {Physics of Semiconductor Devices},
  edition   = {3},
  publisher = {Wiley},
  address   = {Hoboken},
  year      = {2006},
  doi       = {10.1002/0470068329},
  isbn      = {978-0-471-14323-9}
}

@article{Shoji2015,
  author  = {Shoji, Tomoyuki and Nishida, Shuichi and Hamada, Kimimori and Tadano, Hiroshi},
  title   = {Analysis of Neutron-Induced Single-Event Burnout in SiC Power MOSFETs},
  journal = {Microelectronics Reliability},
  volume  = {55},
  number  = {9--10},
  pages   = {1517--1521},
  year    = {2015},
  doi     = {10.1016/j.microrel.2015.06.081}
}

@article{McPherson2020,
  author  = {McPherson, Joseph A. and Hitchcock, Collin W. and Chow, T. Paul and Ji, Wei and Woodworth, Andrew A.},
  title   = {Mechanisms of Heavy Ion-Induced Single Event Burnout in 4H-SiC Power MOSFETs},
  journal = {Materials Science Forum},
  volume  = {1004},
  pages   = {889--896},
  year    = {2020},
  doi     = {10.4028/www.scientific.net/MSF.1004.889}
}

@article{Liao2024,
  author  = {Liao, Qiulan and Liu, Hongxia},
  title   = {Research on Single-Event Burnout Reinforcement Structure of SiC MOSFET},
  journal = {Micromachines},
  volume  = {15},
  number  = {5},
  pages   = {642},
  year    = {2024},
  doi     = {10.3390/mi15050642}
}

@article{Wang2024,
  author  = {Wang, Haibin and Gu, Jianghao and Huang, Xiaofeng and others},
  title   = {Single-Event Burnout Resilient Design of 4H-SiC MOSFETs through Staircase-Like Buffer Layer},
  journal = {Microelectronics Reliability},
  volume  = {154},
  pages   = {115344},
  year    = {2024},
  doi     = {10.1016/j.microrel.2024.115344}
}

@article{Peng2021,
  author  = {Peng, C. and Zhang, X. and Guo, H. and Zhang, F. and Pan, X. and Liu, Y. and Gu, Z. and Ju, A. and Ouyang, X.},
  title   = {Experimental and Simulation Studies of Radiation-Induced Single Event Burnout in SiC-Based Power MOSFETs},
  journal = {IET Power Electronics},
  volume  = {14},
  number  = {9},
  pages   = {1700--1712},
  year    = {2021},
  doi     = {10.1049/pel2.12147}
}

@article{Zwanzig1961,
  author  = {Zwanzig, Robert},
  title   = {Memory Effects in Irreversible Thermodynamics},
  journal = {Physical Review},
  volume  = {124},
  number  = {4},
  pages   = {983--992},
  year    = {1961},
  doi     = {10.1103/PhysRev.124.983}
}

@article{Mori1965,
  author  = {Mori, Hazime},
  title   = {Transport, Collective Motion, and Brownian Motion},
  journal = {Progress of Theoretical Physics},
  volume  = {33},
  number  = {3},
  pages   = {423--455},
  year    = {1965},
  doi     = {10.1143/PTP.33.423}
}

@article{Sexton2003,
  author  = {Sexton, F. W.},
  title   = {Destructive Single-Event Effects in Semiconductor Devices and ICs},
  journal = {IEEE Transactions on Nuclear Science},
  volume  = {50},
  number  = {3},
  pages   = {603--621},
  year    = {2003},
  doi     = {10.1109/TNS.2003.813137}
}

@article{Padovani2024,
  author  = {Padovani, Andrea and La Torraca, Paolo and Strand, Jack and Larcher, Luca and Shluger, Alexander L.},
  title   = {Dielectric breakdown of oxide films in electronic devices},
  journal = {Nature Reviews Materials},
  volume  = {9},
  number  = {9},
  pages   = {607--627},
  year    = {2024},
  doi     = {10.1038/s41578-024-00702-0}
}

@article{Gabriel2023,
  author  = {Gabriel, Okafor Ekene and Huitink, David Ryan},
  title   = {Failure Mechanisms Driven Reliability Models for Power Electronics: A Review},
  journal = {Journal of Electronic Packaging},
  volume  = {145},
  number  = {2},
  pages   = {020801},
  year    = {2023},
  doi     = {10.1115/1.4055774}
}

@article{Zapperi1999,
  author  = {Zapperi, Stefano and Ray, Purusattam and Stanley, H. Eugene and Vespignani, Alessandro},
  title   = {Avalanches in breakdown and fracture processes},
  journal = {Physical Review E},
  volume  = {59},
  number  = {5},
  pages   = {5049--5057},
  year    = {1999},
  doi     = {10.1103/PhysRevE.59.5049}
}

@article{Spang2024,
  author  = {Spang, Arne and Thielmann, Marcel and Kiss, D{\'a}niel},
  title   = {Rapid Ductile Strain Localization Due to Thermal Runaway},
  journal = {Journal of Geophysical Research: Solid Earth},
  volume  = {129},
  number  = {10},
  pages   = {e2024JB028846},
  year    = {2024},
  doi     = {10.1029/2024JB028846}
}

@article{Fu2024,
  author  = {Fu, Jinping and Du, Wei and Hou, Huiming and Zhao, Xiaohua},
  title   = {Avalanche scaling law for heterogeneous interfacial fracture},
  journal = {Physica A: Statistical Mechanics and its Applications},
  volume  = {639},
  pages   = {129682},
  year    = {2024},
  doi     = {10.1016/j.physa.2024.129682}
}

@article{Kuehn2011,
  author  = {Kuehn, Christian},
  title   = {A mathematical framework for critical transitions: Bifurcations, fast--slow systems and stochastic dynamics},
  journal = {Physica D: Nonlinear Phenomena},
  volume  = {240},
  number  = {12},
  pages   = {1020--1035},
  year    = {2011},
  doi     = {10.1016/j.physd.2011.02.012}
}

@article{Gillespie1977,
  author  = {Gillespie, Daniel T.},
  title   = {Exact stochastic simulation of coupled chemical reactions},
  journal = {The Journal of Physical Chemistry},
  volume  = {81},
  number  = {25},
  pages   = {2340--2361},
  year    = {1977},
  doi     = {10.1021/j100540a008}
}

@article{Weller2003,
  author  = {Weller, R. A. and Sternberg, A. L. and Massengill, L. W. and Schrimpf, R. D. and Fleetwood, D. M.},
  title   = {Evaluating average and atypical response in radiation effects simulations},
  journal = {IEEE Transactions on Nuclear Science},
  volume  = {50},
  number  = {6},
  pages   = {2265--2271},
  year    = {2003},
  doi     = {10.1109/TNS.2003.821576}
}

@article{Schrimpf2007,
  author  = {Schrimpf, Ronald D. and Weller, Robert A. and Mendenhall, Marcus H. and Reed, Robert A. and Massengill, Lloyd W.},
  title   = {Physical mechanisms of single-event effects in advanced microelectronics},
  journal = {Nuclear Instruments and Methods in Physics Research Section B: Beam Interactions with Materials and Atoms},
  volume  = {261},
  number  = {1--2},
  pages   = {1133--1136},
  year    = {2007},
  doi     = {10.1016/j.nimb.2007.04.050}
}

@article{Ham2024,
  author  = {Ham, Lucy and Coomer, Megan A. and {\"O}cal, Kaan and Grima, Ramon and Stumpf, Michael P. H.},
  title   = {A stochastic vs deterministic perspective on the timing of cellular events},
  journal = {Nature Communications},
  volume  = {15},
  pages   = {5286},
  year    = {2024},
  doi     = {10.1038/s41467-024-49624-z}
}

@article{Reed2007,
  author  = {Reed, R. A. and Weller, R. A. and Mendenhall, M. H. and Lauenstein, J.-M. and Warren, K. M. and Pellish, J. A. and Schrimpf, R. D. and Sierawski, B. D. and Massengill, L. W. and Dodd, P. E. and Shaneyfelt, M. R. and Felix, J. A. and Schwank, J. R. and Haddad, N. F. and Lawrence, R. K. and Bowman, J. H. and Conde, R.},
  title   = {Impact of Ion Energy and Species on Single Event Effects Analysis},
  journal = {IEEE Transactions on Nuclear Science},
  volume  = {54},
  number  = {6},
  pages   = {2312--2321},
  year    = {2007},
  doi     = {10.1109/TNS.2007.909901}
}

@article{Redner2023,
  author  = {Redner, S.},
  title   = {A first look at first-passage processes},
  journal = {Physica A: Statistical Mechanics and its Applications},
  volume  = {631},
  pages   = {128545},
  year    = {2023},
  doi     = {10.1016/j.physa.2023.128545}
}

@article{Schuss2007,
  author  = {Schuss, Z. and Singer, A. and Holcman, D.},
  title   = {The narrow escape problem for diffusion in cellular microdomains},
  journal = {Proceedings of the National Academy of Sciences},
  volume  = {104},
  number  = {41},
  pages   = {16098--16103},
  year    = {2007},
  doi     = {10.1073/pnas.0706599104}
}

@article{Grome2024,
  author  = {Grome, Christopher A. and Ji, Wei},
  title   = {A Brief Review of Single-Event Burnout Failure Mechanisms and Design Tolerances of Silicon Carbide Power MOSFETs},
  journal = {Electronics},
  volume  = {13},
  number  = {8},
  pages   = {1414},
  year    = {2024},
  doi     = {10.3390/electronics13081414}
}

@article{Li2025,
  author  = {Li, Ming and Yang, Ru and Chen, Xiaoqing and Liu, Xiaoyan and Wang, Qiang and Liu, Jie and Wang, Yuming},
  title   = {Single event burnout in SiC MOSFETs induced by nuclear reactions with high-energy oxygen ions},
  journal = {Chinese Physics B},
  year    = {2025},
  doi     = {10.1088/1674-1056/adcd45}
}

@article{Mo2023,
  author  = {Mo, Lihua and Yu, Quanzhi and Hu, Zhiliang and Zhou, Bin and Yi, Tiancheng and Yuan, Liubin and Lin, Li and Shen, Fei and Liang, Tianjiao},
  title   = {Single event burnout of SiC MOSFET induced by atmospheric neutrons},
  journal = {Microelectronics Reliability},
  volume  = {146},
  pages   = {114997},
  year    = {2023},
  doi     = {10.1016/j.microrel.2023.114997}
}

@article{Pocaterra2023,
  author  = {Pocaterra, Marco and Ciappa, Mauro},
  title   = {Single event burnout failures caused in silicon carbide power devices by alpha particles emitted from radionuclides},
  journal = {e-Prime: Advances in Electrical Engineering, Electronics and Energy},
  volume  = {5},
  pages   = {100203},
  year    = {2023},
  doi     = {10.1016/j.prime.2023.100203}
}

@article{Kuboyama1992,
  author  = {Kuboyama, S. and Matsuda, S. and Kanno, T. and Ishii, T.},
  title   = {Mechanism for single-event burnout of power MOSFET's and its characterization technique},
  journal = {IEEE Transactions on Nuclear Science},
  volume  = {39},
  number  = {6},
  pages   = {1698--1703},
  year    = {1992},
  doi     = {10.1109/23.211356}
}

@article{WangNegGate2024,
  author  = {Wang, Haibin and Nie, Zhichao and Huang, Xiaofeng and Gu, Jianghao and Tan, Zhixin and Jing, Hantao and Mo, Lihua and Hu, Zhiliang and Wang, Xueming},
  title   = {The impact of negative gate voltage on neutron-induced single event effects for SiC MOSFETs},
  journal = {Microelectronics Reliability},
  volume  = {163},
  pages   = {115547},
  year    = {2024},
  doi     = {10.1016/j.microrel.2024.115547}
}

@article{Germanicus2025,
  author  = {Coq Germanicus, Rosine and Michez, Alain and Niskanen, Kimmo and Chaudhary, Mahima and Bascoul, Guillaume and Chazal, Vanessa and Wrobel, Frederic and Boch, Jerome},
  title   = {Single Event Effects of SiC Power MOSFETs: From Neutron Interaction to Destruction at the Die Level},
  journal = {IEEE Transactions on Nuclear Science},
  volume  = {72},
  number  = {8},
  pages   = {2368--2376},
  year    = {2025},
  doi     = {10.1109/TNS.2025.3561583}
}

@article{Yuan2024,
  author  = {Yuan, Zimo and Lim, Jang-Kwon and Metreveli, Alex and Krishna Murthy, Hithiksha and Bakowski, Mietek and Hall{\'e}n, Anders},
  title   = {Single Event Effects in 3.3 kV 4H-SiC MOSFETs Due to MeV Ion Impact},
  journal = {Solid State Phenomena},
  volume  = {361},
  pages   = {77--83},
  year    = {2024},
  doi     = {10.4028/p-90Xrjk}
}

@article{Martinella2023,
  author  = {Martinella, Corinna and Race, Salvatore and Stark, Roger and Garc{\'i}a Al{\'i}a, Ruben and Javanainen, Arto and Grossner, Ulrike},
  title   = {High-Energy Proton and Atmospheric-Neutron Irradiations of SiC Power MOSFETs: SEB Study and Impact on Channel and Drift Resistances},
  journal = {IEEE Transactions on Nuclear Science},
  volume  = {70},
  number  = {8},
  pages   = {1844--1851},
  year    = {2023},
  doi     = {10.1109/TNS.2023.3267144}
}

@article{Wang2019,
  author  = {Wang, Ying and Lin, Mao and Li, Xing Ji and Wu, Xue and Yang, Jian Qun and Bao, Meng Tian and Yu, Cheng Hao and Cao, Fei},
  title   = {Single-Event Burnout Hardness for the 4H-SiC Trench-Gate MOSFETs Based on the Multi-Island Buffer Layer},
  journal = {IEEE Transactions on Electron Devices},
  volume  = {66},
  number  = {10},
  pages   = {4264--4272},
  year    = {2019},
  doi     = {10.1109/TED.2019.2933026}
}

@article{Bi2020,
  author  = {Bi, Jian-Xiong and Wang, Ying and Wu, Xue and Li, Xing-Ji and Yang, Jian-Qun and Bao, Meng-Tian and Cao, Fei},
  title   = {Single-Event Burnout Hardening Method and Evaluation in SiC Power MOSFET Devices},
  journal = {IEEE Transactions on Electron Devices},
  volume  = {67},
  number  = {10},
  pages   = {4340--4345},
  year    = {2020},
  doi     = {10.1109/TED.2020.3015718}
}

@article{Sun2025,
  author  = {Sun, Shuqing and Chen, Feida and Sun, Yongbo and Li, Yongxing and Yang, Kun and Tang, Xiaobin},
  title   = {Single Event Effects Hardening in SiC Double-Trench MOSFETs},
  journal = {Microelectronics Reliability},
  volume  = {164},
  pages   = {115569},
  year    = {2025},
  doi     = {10.1016/j.microrel.2024.115569}
}

@article{Alberton2022,
  author  = {Alberton, Saulo G. and Aguiar, V. A. P. and Medina, N. H. and Added, N. and Macchione, E. L. A. and Menegasso, R. and Ces{\'a}rio, G. J. and Santos, H. C. and Scarduelli, V. B. and Alc{\'a}ntara-N{\'u}{\~n}ez, J. A. and Guazzelli, M. A. and Santos, R. B. B. and Flechas, D.},
  title   = {Charge deposition analysis of heavy-ion-induced single-event burnout in low-voltage power VDMOSFET},
  journal = {Microelectronics Reliability},
  volume  = {137},
  pages   = {114784},
  year    = {2022},
  doi     = {10.1016/j.microrel.2022.114784}
}

@article{Hanggi1990,
  title = {Reaction-rate theory: fifty years after Kramers},
  author = {H\"anggi, Peter and Talkner, Peter and Borkovec, Michal},
  journal = {Rev. Mod. Phys.},
  volume = {62},
  issue = {2},
  pages = {251--341},
  numpages = {0},
  year = {1990},
  month = {Apr},
  publisher = {American Physical Society},
  doi = {10.1103/RevModPhys.62.251}
}

@article{Sengupta2024,
  author  = {Sengupta, Arijit and Ball, Dennis R. and Sternberg, Andrew L. and Islam, Sajal and Senarath, Aditha S. and Reed, Robert A. and McCurdy, Michael W. and Zhang, En Xia and Hutson, John M. and Alles, Michael L. and Osheroff, Jason M. and Jacob, Biju and Hitchcock, Collin W. and Goswami, Shubhodeep and Schrimpf, Ronald D. and Galloway, Kenneth F. and Witulski, Arthur F.},
  title   = {LET and Voltage Dependence of Single-Event Burnout and Single-Event Leakage Current in High-Voltage SiC Power Devices},
  journal = {IEEE Transactions on Nuclear Science},
  volume  = {71},
  number  = {4},
  pages   = {809--815},
  year    = {2024},
  doi     = {10.1109/TNS.2024.3357129}
}

@article{Ziegler2010,
  author  = {Ziegler, James F. and Ziegler, M. D. and Biersack, J. P.},
  title   = {{SRIM} -- The stopping and range of ions in matter (2010)},
  journal = {Nuclear Instruments and Methods in Physics Research Section B: Beam Interactions with Materials and Atoms},
  volume  = {268},
  number  = {11--12},
  pages   = {1818--1823},
  year    = {2010},
  doi     = {10.1016/j.nimb.2010.02.091}
}

@article{Agostinelli2003,
  author  = {Agostinelli, S. and et al},
  title   = {{GEANT4} -- A simulation toolkit},
  journal = {Nuclear Instruments and Methods in Physics Research Section A: Accelerators, Spectrometers, Detectors and Associated Equipment},
  volume  = {506},
  number  = {3},
  pages   = {250--303},
  year    = {2003},
  doi     = {10.1016/S0168-9002(03)01368-8}
}

\end{document}